\documentclass[10pt,journal,cspaper,compsoc]{IEEEtran} 
\usepackage{cite}
\usepackage{pslatex}
\usepackage[pdftex]{graphicx}
\usepackage[caption=false,font=footnotesize]{subfig}
\usepackage{fixltx2e}
\usepackage{stfloats}
\usepackage{url}
\usepackage{color}
\usepackage{multicol}
\usepackage{balance}
\usepackage{amssymb}
\usepackage[lined,linesnumbered]{algorithm2e}

\newtheorem{definition}{Definition}
\begin{document}
\title{FASHION: Fault-Aware Self-Healing Intelligent On-chip Network} 
\author{
Pengju Ren,~\IEEEmembership{Member ~IEEE},
Michel A. Kinsy,~\IEEEmembership{Member ~IEEE},
Mengjiao Zhu,~
Shreeya Khadka, Mihailo Isakov, \\
Aniruddh Ramrakhyani,~
Tushar Krishna~\IEEEmembership{Member ~IEEE}
and Nanning Zheng,~\IEEEmembership{Fellow ~IEEE}

\IEEEcompsocitemizethanks{
\IEEEcompsocthanksitem Pengju Ren, Mengjiao Zhu and Nanning Zheng are with the Xi'an Jiaotong University, Xi'an, Shaanxi, P.R.China, 710049. E-mail: pengjuren@mail.xjtu.edu, nnzheng@mail.xjtu.edu.cn.
\IEEEcompsocthanksitem Michel Kinsy, Shreeya Khadka and Mihailo Isakov are with the Department of Electrical and Computer Engineering at the Boston University, MA 02215, U.S.A. E-mail:mkinsy@bu.edu.
\IEEEcompsocthanksitem Aniruddh Ramrakhyani and Tushar Krishna are with the School of Electrical and Computer Engineering at the Georgia Institute of Technology, GA 30332, U.S.A. E-mail:tushar@ece.gatech.edu
}
\thanks{}}

\IEEEcompsoctitleabstractindextext{
\begin{abstract}
To avoid packet loss and deadlock scenarios that arise due to faults or power gating in multicore and many-core systems, 
the network-on-chip needs to possess resilient communication and load-balancing properties.
In this work, we introduce the \textbf{\textit{Fashion}} router, a self-monitoring and self-reconfiguring design 
that allows for the on-chip network to dynamically adapt to component failures. 
First, we introduce a distributed intelligence unit, called \textit{Self-Awareness Module (SAM)}, 
which allows the router to detect permanent component failures and build a network connectivity map. 
Using local information, \textit{SAM} adapts to faults, guarantees connectivity and deadlock-free routing inside the \textit{maximal connected subgraph} and keeps routing tables up-to-date. 
Next, to reconfigure network links or virtual channels around faulty/power-gated components, we add bidirectional link and unified virtual channel structure features to the \textbf{\textit{Fashion}} router. 
This version of the router, named \textbf{\textit{Ex-Fashion}}, further mitigates the negative system performance impacts, leads to larger \textit{maximal connected subgraph} and sustains a relatively high degree of fault-tolerance.
To support the router, we develop a fault diagnosis and recovery algorithm executed by the Built-In Self-Test, self-monitoring, 
and self-reconfiguration units at runtime to provide fault-tolerant system functionalities.
The \textbf{\textit{Fashion}} router places no restriction on topology, position or number of faults. It drops 
54.3$\sim$55.4\% fewer nodes for same number of faults (between 30 and 60 faults) in an 8x8 2D-mesh 
over other state-of-the-art solutions. It is scalable and efficient. The area overheads are 2.311\% and 2.659\% when implemented in 8x8 and 16x16 2D-meshes using the \textsc{TSMC} 65nm library at 1.38GHz clock frequency. 
\end{abstract}
}
\maketitle

\section{Introduction}
\label{sec:intro}
With advances in semiconductor technology, continued scaling and integration of transistors have allowed more system functions to be implemented on chip. The current state-of-the-art in computer architecture design is multicore and many-core systems. A growing number 
of these systems is being built with 48\cite{howard201048}, 64\cite{bell_tile64_2008}, 80\cite{intel80_ISSCC_2007}, 100\cite{ramey2011tile}, 256\cite{chen201416} and 4096\cite{merolla2014million} cores. 
In these architectures, processing cores are connected together using a variety of fabric interconnect technologies. Network-on-chip 
(NoC) has emerged as the \textit{de facto} communication fabric in large multicore and many-core architectures\cite{jerger2009chip}\cite{radu:2009}, primarily due to the lack of scalability associated with bus-based communication infrastructures\cite{Ivanov:2005:TNO}. 
The modular structure of NoCs supports more concurrent communications and makes them more flexible for different application 
communication demands, quality of service (QoS) guarantees and resiliency constraints\cite{dally2004principles}. 

As technology scales to the 10-nm regime, the extreme shrinking of transistor feature size and diminishing supply voltage 
are making circuits more sensitive to manufacturing process and environmental variations\cite{borkar2005designing}\cite{reddi2011resilient}. These new semiconductor devices are highly susceptible to early manufacturing defects, and also to latent 
defects because of time-dependent degradation and material wear-out, such as oxide breakdown, negative-bias-temperatur-instability (NBTI), \\channel-hot-carrier (CHC) and time-dependent-dielectric-\\breakdown (TDDB)\cite{wilson2013international}. Positive-bias-temperature-instability (PBTI) on NMOS has also become more pronounced because the adoption of high-\textit{k} gate dielectric\cite{Yu2011Modeling}.  

The ability of chips to self-monitor 
and to self-reconfigure in the presence of faults throughout their lifetime is an important design paradigm going forward\cite{collet2011chip}.
Dynamic 
hardware-supported fault-detection algorithms are being implemented \cite{SastryHari:2009}. Systems, where architecture states and 
systems parameters are periodically copied and stored to allow rollbacks with built-in restoration mechanisms, have been 
proposed\cite{Prvulovic}.
Beyond hardware 
solutions, software approaches that can be used to complement hardware functions and further mitigate 
fault-induced system performance impacts have also been suggested\cite{borkar2005challenges}\cite{sylvester2006elastic}\cite{deorio2011drain}\cite{radu:2009}. 

Unfortunately, the Network-on-Chip (NoC) layer of the architecture is not immune to these 
effects. In fact, since the NoC provides connectivity between the various components of the system, 
fault-tolerance has 
become an essential feature that must be brought to the forefront of the NoC design. 
It is important to maximize the on-chip resource utilization throughout the lifetime of multicore and manycore systems when the reduction of component reliability is unavoidable. 
Router microarchitectures and routing techniques that are more resilient 
to component failures and guarantee a higher degree of node connectivity have started 
being investigated\cite{Kim07faulttolerant, Dongkook, parikh2013udirec}. 
Router/link faults in NoCs change the underlying topology. This introduces three key challenge for fault-tolerance: (1) Deadlocks: The disconnection of certain routers and links can lead to cyclic dependencies 
between network resources, leading to a network deadlock. To avoid deadlocks, cyclic dependencies in the channel dependency graph of each new topology 
need to be removed.
(2) Performance: Faults lead to a loss of path diversity in the NoC, and higher congestion for certain paths. Adaptive routing\cite{shafiee2011application}\cite{fu2011abacus} can mitigate some of these effects, but have to remove certain channel dependencies or 
prohibit certain routing paths, which reduces routing path diversity even further\cite{Ren2015ftbr}. (3) Scalability: As we add more cores in a system, 
the size of the NoC grows and warrants a distributed solution since centralized solutions are not going to scale.



The key contribution of this work is a holistic and scalable NoC reliability solution called Fault-Aware Self-Healing Intelligent On-chip Network (FASHION) that address all three challenges listed above. We introduce a distributed intelligence unit called 
\textit{Self-Awareness Module (SAM)}, that allows the router to (1) automatically detect permanent component failures and generate a network connectivity graph through a distributed spanning tree search algorithm with computational complexity of $O(|L|)$, where $L$ is the link number in the network, (2) implement in-hardware self-adjusting techniques to guarantee connected and deadlock-free routes inside the \textit{maximal connected subgraph} with computational complexity of $O(|R||L|)$, where $R$ is the number of nodes in a network, and (3) support the adoption of bidirectional links and unified virtual channel structure to further strengthen the network connectivity and sustain a relatively high degree of fault-tolerance. This work merges these concepts together to provide a robust and practical solution for mitigating fault-related system performance degradation.


%
%
FASHION can also be extended to the NoC power gating domain where links and routers are power-gated to reduce NoC static power dissipation.
  Several previous works \cite{nord, router_parking, panthre} have come up with innovative power-gating techniques to selectively power-gate low utilization NoC components while providing minimal disruption to the NoC traffic. Power-gating of links and routers, however, creates deadlock-prone irregular topologies that change dynamically, thereby introducing the same performance and deadlock challenges listed earlier. This work proposes a new re-configuration algorithm that provides deadlock-free paths in an arbitrary irregular topology and can thus be leveraged by existing works in NoC power-gating domain to further strengthen their schemes.
%
%

The rest of the paper is organized as follows. Section \ref{sec:related} highlights the related works. The effects of faults on network connectivity in NoC-based architecture are introduced in Section \ref{sec:connectivity}. Router micro-architecture details are presented in Section \ref{sec:arch}. Section \ref{sec:algo} describes the \textbf{\textit{Fashion}} fault detection and reconfiguration algorithm. Section \ref{sec:ex_fashion} presents the extended \textbf{\textit{Fashion}} architecture. Evaluations and discussions are presented in Section \ref{sec:eval}. Finally, Section \ref{sec:concl} concludes the paper. 

\section{Related Work}
\label{sec:related}
Modern computer architectures already implement a variety of mechanisms to improve system reliability. Existing approaches for NoC resilience are either on-line or off-line. Off-line reconfiguration algorithms\cite{gomez2006routing}\cite{fukushima2009fault}\cite{ren2014fault} use knowledge of the underlying network topology to compute routing decisions. On-line solutions are implemented in hardware under a tight area and power budget and can leverage the on-chip resources to run both detection and recovery programs. On-line solutions are more attractive than off-line solutions, because they generally consume fewer resources and have better scalability and computational efficiency. The \textbf{\textit{Fashion}} architecture uses an on-line 
fault-tolerance approach.  

A large number of fault-tolerant routing algorithms have also been proposed.
Immunet\cite{puente2004immunet} is able to tolerate any combination of faults as long as the network is connected. Fick {\it et al.} introduced a low overhead routing algorithm named Vicis\cite{fick2009vicis}, Gomez {\it et al.} provided an intermediate node based fault-tolerant routing by using escape virtual channels in each phase\cite{gomez2006routing}, Ariadne\cite{aisopos2011ariadne} leveraged Up*/Down* routing\cite{schroeder1991autonet}, which assigned each link either ``up'' or ``down'' direction and disallowed transmission from ``down'' to ``up'' to break cyclic dependencies. A distributed and lightweight fault-tolerant routing algorithm, named Hermes\cite{iordanou2014hermes} was proposed by Iordanou {\it et al.}. It applies XY- or O1TURN-routing for high throughput, while choses the Up*/Down* routing rule to ensure acyclic path formations to avoid deadlock and achieve fault-tolerance. 

Sylvester {\it et al.}\cite{sylvester2006elastic} claimed that an on-line \textit{self-organization mechanism} is crucially important to improve the effective yield throughout the lifetime of SoCs. The ForEVeR algorithm\cite{parikh2011formally}, applies formal methods and runtime verification to ensure functional correctness in NoCs. Murali {\it et al.} demonstrate that hybrid error detection and correction mechanisms provide better performance\cite{murali2005analysis}. A recovery mechanism named \textsc{drain} was introduced in\cite{deorio2011drain} to provide system-level recovery for any number of disconnected nodes caused by permanent failures by sending the architectural state and dirty cached data from disconnected nodes to healthy caches nearby to reduce the effects of failures. 

Other architecture level improvements for NoC reliability have been studied. Kim {\it et al.} proposed a row-column decoupled router\cite{kim2006gracefully}, which employed decoupled parallel arbiters and small crossbars. Palesi {\it et al.} introduced a scheme to efficiently use partially-faulty links\cite{palesi2010leveraging}. NoCAlert\cite{prodromou2012nocalert} is an on-line and real-time fault detection mechanism which operates seamlessly and concurrently with normal NoC operation.

Recently, Parikh {\it et al.} proposed a fine-resolution detection and reconfiguration strategy to cope with permanent faults, named uDIREC in\cite{parikh2013udirec}. 
Similar to prior work by  Shamshiri {\it et al.}\cite{shamshiri2011end}, uDIREC uses a ``supervisor'' node to make detection decisions and stores the topology information in a software-maintained scoreboard at that node. It then applies Up*/Down* routing\cite{schroeder1991autonet} to guarantee deadlock avoidance. In Up*/Down* routing, all links are tagged as Up or Down relative to a root node, and a message is not allowed to make Down to Up turns.
In order to maximize network connectivity, uDIREC needs to implement the \textit{breath-first search} in software and discover the optimal solution via an exhaustive search of the ``root'' node. 
We find that the Up*/Down* routing scheme disables a large percentage of turns to break cycles which lower routing options and sacrifice network performance, because there are only two type of turns (``up-to-down'' and ''down-to-up'').  
In contrast, compared with uDIREC, \textbf{\textit{Fashion}} only need to be executed once to detect the connectivity of the underlying network. Also, our experimental results show that the 
\textbf{\textit{Fashion}} architecture is more efficient in terms of computation complexity, scalability and the degree of fault tolerance.
According to our experiments (see Section~\ref{sec:eval}), the average percentage of forbidden turns of \textbf{\textit{Fashion}} is 14.1\% less than Up*/Down* routing.


In the NoC power-gating domain, NorD~\cite{nord} a recent work, uses a high-latency ring snaking through the network as the escape path. Packets are routed adaptively till their mis-routed hop-count increases beyond a certain threshold upon which they are forced to enter the escape path. Router-parking~\cite{router_parking} replaces the high latency ring of NorD with an escape-path constructed using up-down routing. Deadlocks in this scheme are detected using counters and one Virtual Channel per Virtual Network per Input port is always kept reserved for the escape path. Panthre~\cite{panthre} does away with the escape-path and instead uses Up*/Down* routing for all the VCs. However, as pointed out earlier, Up*/Down* routing provides deadlock-freedom at the cost of significant reduction in path diversity. In addition, certain traffic flows are forced to use non-minimal paths leading to increased latency and energy consumption. In contrast, FASHION provides 4.80\%, 6.06\% and 14.67\% lower latency to traffic on average compared to up-down routing at 5, 10 and 15 power-gated links respectively. Moreover, it provides low re-configuration time which is very useful to schemes in NoC power-gating domain as power-gating decisions are taken every epoch (usually 10K cycles), thus requiring the network to be re-configured every epoch if there is a change in the active topology.

\section{\textbf{Effects of Faults on Network Connectivity in NoC-based Architectures}}
\label{sec:connectivity}
For the communication infrastructure to dynamically identify components with permanent or transient faults and to 
autonomously self-reconfigure, the network router needs to exhibit some inner intelligence. 
To describe the \textbf{\textit{Fashion}} architecture and its functionality, we begin by giving the standard definitions for 
\textit{network channel graphs} and \textit{cut-elements}. 
\begin{definition}
\textit{Given a network-on-chip characterization graph $G =G(R,L)$, where the routers and links in the network are 
given by the sets $R$ and $L$, $r_i \in R$ represents  the router associated with processor element $i$, while each 
arc $l_{i,j} \in L$ represents a link from $r_i$ to $r_j$.}
For a given $G$, two vertices $i$ and $j$ are called \textbf{connected elements} if $G$ contains a path from $i$ to $j$; 
graph $G$ is said to be \textbf{connected} if every pair of vertices in $G$ is connected; A \textbf{cut vertex} of $G$ is a 
vertex whose removal results in a disconnected $G$; A \textbf{cut edge} of $G$ is an arc, whose removal disconnects 
$G$. Together \textbf{cut vertices} and \textbf{cut edges} form \textbf{cut elements} and are critical to the connectivity 
of $G$. 
\end{definition}

\begin{definition}
A \textbf{rooted acyclic graph} is a graph in which one of the vertices is distinguished from the others. This particular vertex is called the \textbf{root} of the graph. In a rooted graph $G$ with root $r_{root}$, any node $j$ on the unique path from $r_{root}$ to a node $i$ is called an \textbf{ancestor} of $i$, and $i$ is called a \textbf{descendant} of $j$. If the last edge on the path from the root $r_{root}$ of the graph $G$ to a node $i$ is $(j, i)$, the $j$ is the \textbf{parent} of $i$, and $i$ is called a \textbf{child} of $j$, denoted $i = j_{child}$ and $j = i_{parent}$. If two nodes have the same parent, they are \textbf{siblings}; a node with no children is a \textbf{leaf}. 
\end{definition}

\begin{figure}[t]
\begin{center}
\includegraphics[width=0.35\textwidth]{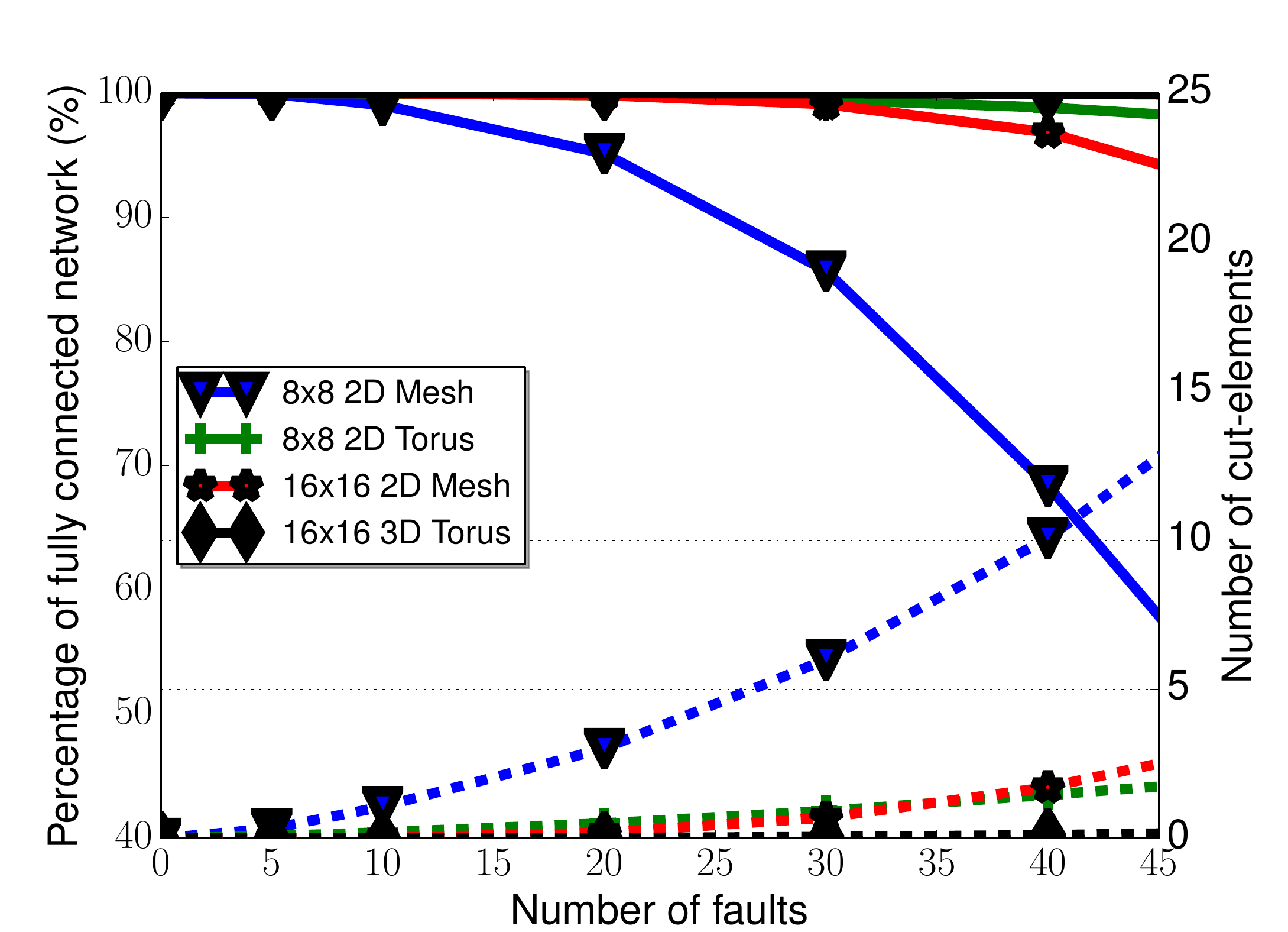}
\caption{Relationship of number of permanent faults and network connectivity and number of \textit{cut elements}. 2D-Meshes with uniform-random faults distribution and the ratio of fault links and fault nodes is 24:1\cite{parikh2013udirec}, the dotted curves represent the number of \textit{cut-elements}.}
\label{fig:faults}
\end{center}
\vspace{-0.2in}
\end{figure}

As stated in the \textit{Definition 1}, a \textit{cut-element} is an element whose removal breaks network connectivity 
and an element may be a \textit{vertex} or an \textit{edge}. The connectivity of a network graph $G$ dictates path diversity 
and routing choices in the NoC. Node pairs can only communicate if they belong to the same 
\textbf{connected subgraph}. The vertex set $V^{cg} = \{G_1, G_2, ..., G_k\}$ contains all the \textit{connected subgraph} $G_i$ of $G$, the \textbf{maximal connected subgraph} $G^{max}$ is the one with the maximal number of vertices. The $V^{cg}$ is completely determined by the topology of the G.  

\setcounter{figure}{2}
\begin{figure*}[b]
\begin{center}
\includegraphics[width=0.9\textwidth]{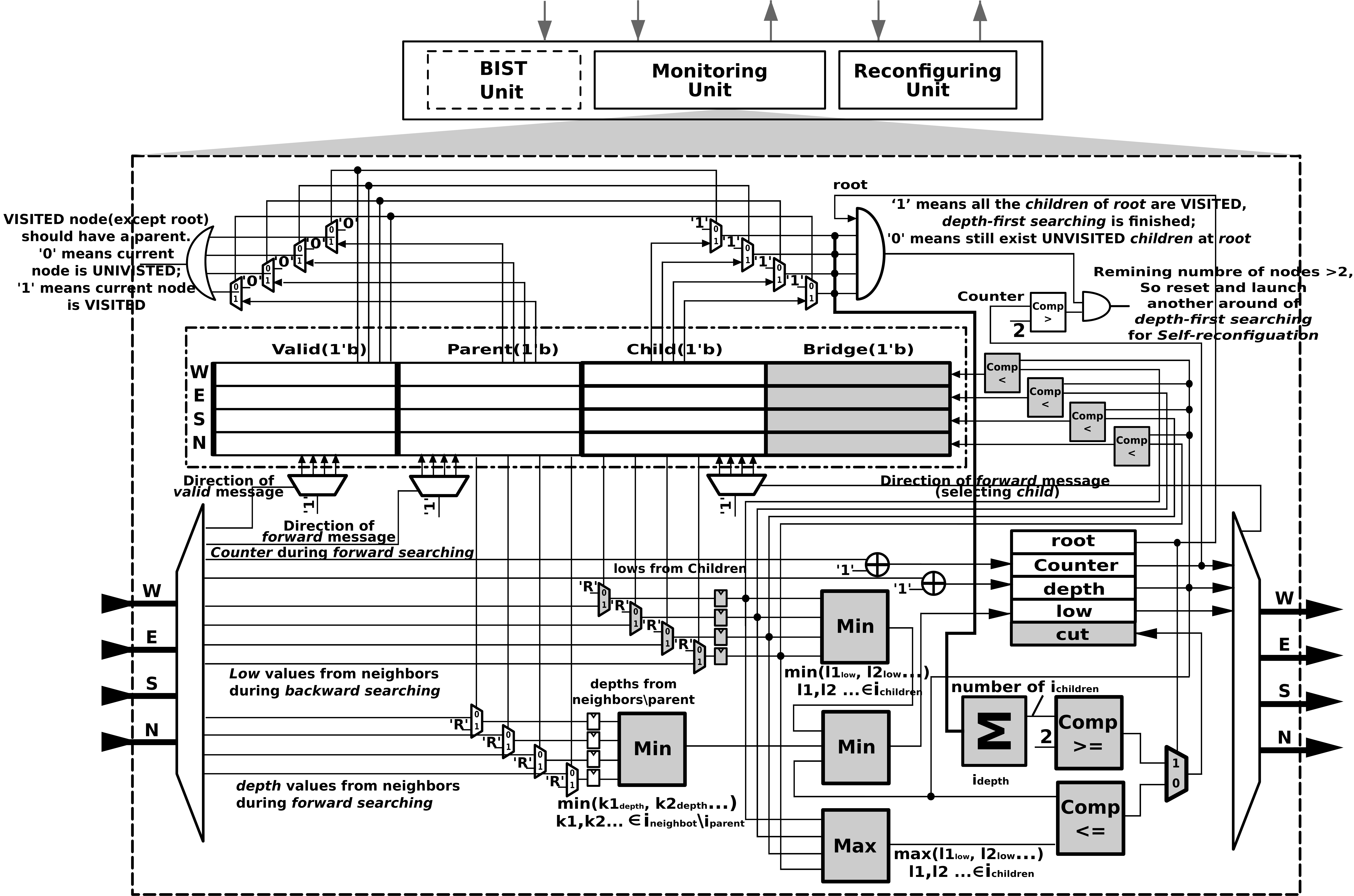}
\caption{Micro-architecture for the \textit{Self-Monitoring Unit}.}
\label{fig:monitor}
\end{center}
\vspace{-0.2in}
\end{figure*}

uDIREC\cite{parikh2013udirec} performed a study where stuck-at faults were injected in a 5-ports wormhole router, in a spatial distribution proportional to the silicon area of gates and wires.
Their analysis revealed that 96\% of faults affect only a small fraction of the router logic. Furthermore, the effects of these faults can be entirely masked by disabling and re-routing around a single router link (e.g. error in the input buffer). The other 4\% of faults can cause the entire router to fail. These faults often involve the router's control logic or critical components like the arbiter or crossbar. Therefore, in the simulation results presented in this paper, we assume a uniform-random distribution of faults with the ratio of faulty links to faulty nodes as 24:1.

Fig~\ref{fig:faults} shows the connectivity impacts of different percentages of faulty components/routers for different sized networks. 
Unpredictable faults may occur at any place in the network, thus we assume a uniform-random distribution of faults over silicon area.
The results presented are the average outcome from 100,000 simulations for each case using the \textsc{hornet}\cite{ren2012hornet} simulator.
For a medium sized 64-node 2D-Mesh network, there are roughly 6 \textit{cut element}s out of 30 faults. This corresponds approximately to 14.73\% of network connectivity loss. Fig~\ref{fig:faults} highlights the relationship between average number of \textit{cut elements} and the percentage of fully connected network. Notice that the \textit{maximal connected subgraph} ($G^{max}$) normally forms an irregular graph, which cannot guarantee all the node-pairs in the $G^{max}$ are able to communicate with each other, because the practical connectivity is determined both by the topology of $G^{max}$ as well as the routing algorithms\cite{Ren2015CBCG}. We will show that \textbf{\textit{Fashion}} ensures connectivity inside the $G^{max}$ in Section~\ref{sec:SR_alg}. 

\setcounter{figure}{1}
\begin{figure}[t]
\begin{center}
\includegraphics[width=0.40\textwidth]{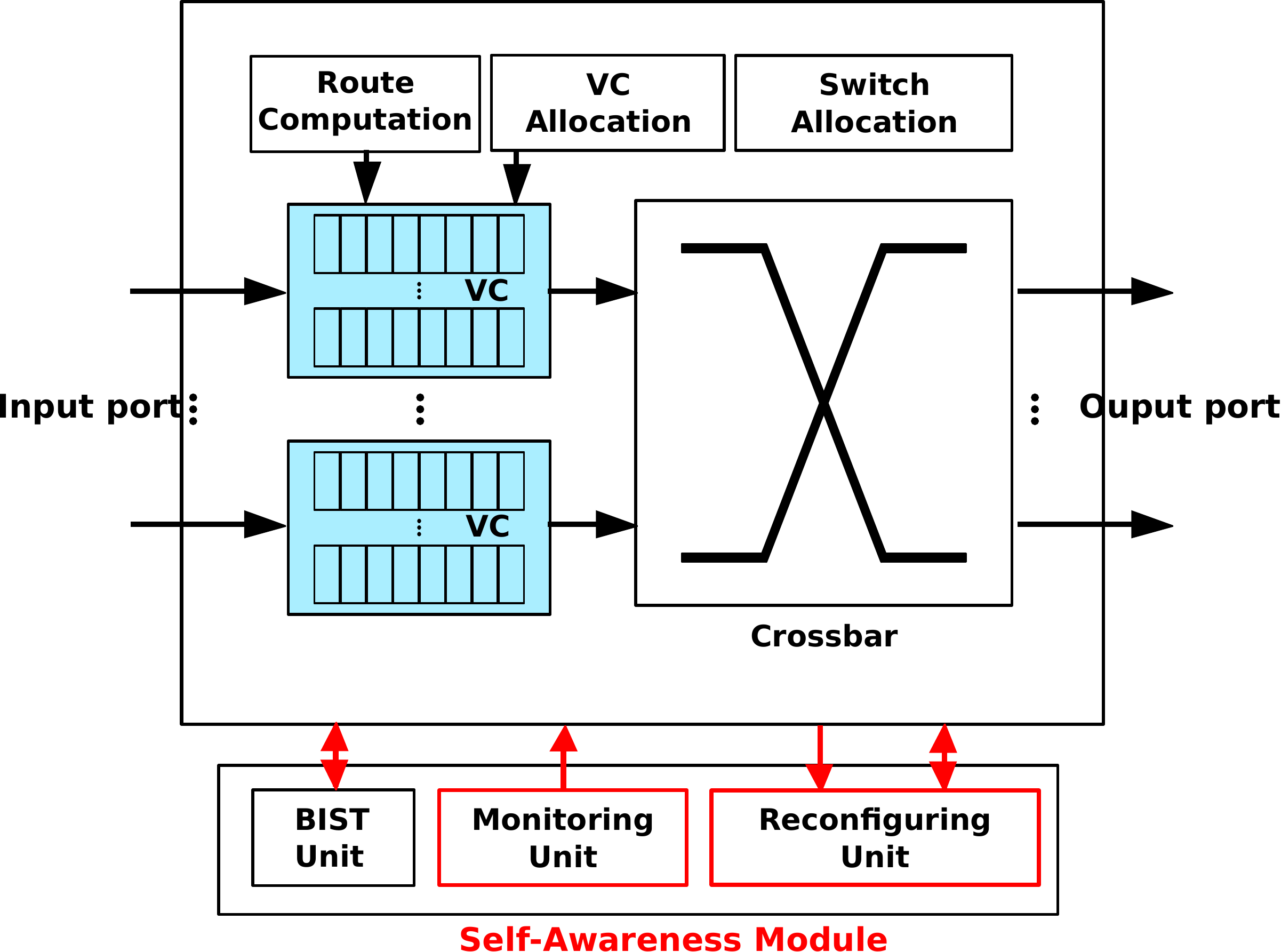}
\caption{\textbf{\textit{Fashion}}: \textbf{F}ault-\textbf{A}ware \textbf{S}elf-\textbf{H}ealing \textbf{I}ntelligent \textbf{O}n-chip \textbf{N}etwork router. }
\label{fig:fashion_router}
\end{center}
\vspace{-0.2in}
\end{figure}

Network connectivity decreases at a faster rate with an increasing number of faults. The need to maximize on-chip resource 
utilization becomes accentuated with decreasing network component reliability. Hence, (1) determining whether defective components are \textit{cut elements} of the network and identifying \textit{connected subgraphs} of a disconnected NoC and (2) mitigating their negative impact on network performance by reconfiguring the NoC and maintaining network connectivity are at the core of \textbf{\textit{Fashion}}'s fault-tolerance scheme.
`
\section{Fashion Router Architecture}
\label{sec:arch}

\subsection{Micro-Architecture of Self-Awareness Module}
In conventional wormhole virtual-channel routers\cite{Dally:2003}, the routing operation generally takes four steps; 
namely, routing (RC), virtual-channel allocation (VA), switch allocation (SA), and switch traversal (ST), 
where each phase corresponds to a pipeline stage in the router. 
In \textbf{\textit{Fashion}}, the architecture is augmented to support runtime self-monitoring 
and self-reconfiguration. Figure~\ref{fig:fashion_router} depicts the \textbf{\textit{Fashion}} architecture. The key 
architectural modification is the introduction of the \textit{Self-Awareness Module (SAM)}. The module contains 
a Built-In Self-Test (BIST) unit, a \textit{Self-Monitoring} unit and a \textit{Self-Reconfiguration} unit. Network \textit{SAM}s also 
generate network probing traffic. 
\textit{SAM} is out of the critical path of the NoC router, each unit is active only during its specific working period and disabled with power-gating in other cases to reduce power and wear-out faults.

\subsection{\textbf{Self-Monitoring unit}}
The \textit{Self-Monitoring} module monitors and captures on-chip network events. It has an embedded \textit{neighbor list table} 
used to record the states of its immediate neighbors and its relationship with them. Each entry in the table corresponds to a port, for a 2D-Mesh/Torus 
network, there are four entries in the table. The $bridge$ entry is set to ``1'' if the link connecting the port to a 
neighbor is detected as a $\textit{cut edge}$, cf., \textsc{lemma 2} in section~\ref{sec:sda}. The $valid$ entry indicates whether or not a neighbor exists or is temporally out-of-service (e.g. due to power-gating)
or permanently removed from the network. Figure~\ref{fig:monitor} shows the \textit{Self-Monitoring} micro-architecture. 

\setcounter{figure}{3}
\begin{figure}[t]
\begin{center}
\includegraphics[width=0.8\linewidth]{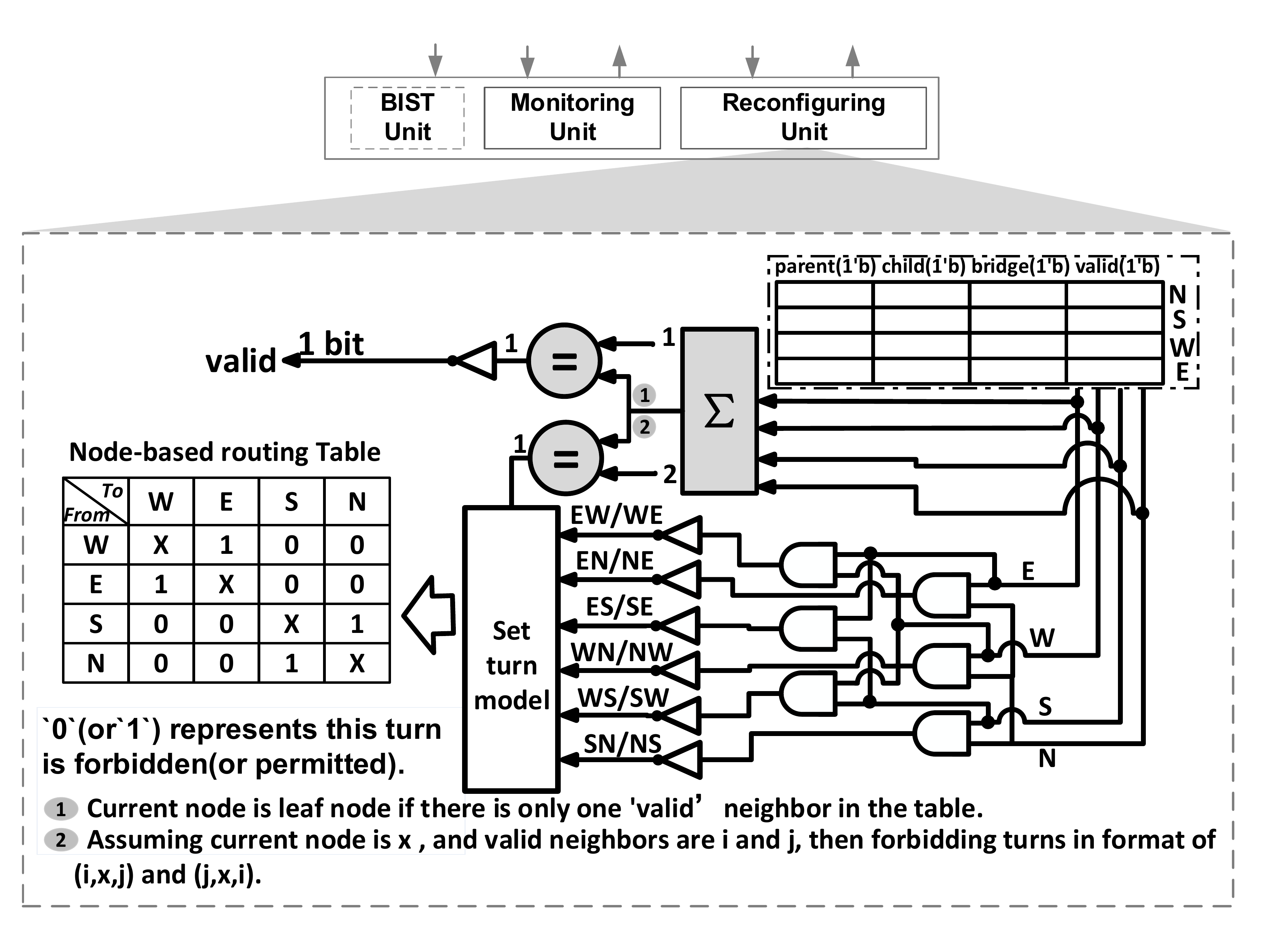}
\caption{\textit{Self-Reconfiguring Unit} overview.}
\label{fig:reconfiguration}
\end{center}
\vspace{-0.2in}
\end{figure}

The \textit{Self-Monitoring} module initiates periodic runs of the \textit{BIST} module by built-in timer. It can also be launched by OS to obtain connectivity information before assigning new applications (or threads) to available processing elements. When the \textit{BIST} unit detects 
 a fault, the \textit{Self-Monitoring} unit is responsible for determining the criticality of the affected component in the network. 
 It does this by building a network connectivity map and labeling all of the \textit{cut-elements} through a fully distributed 
 \textit{depth-first search algorithm}. The details of the algorithm are presented in Section~\ref{sec:sda}.  
It has a one-bit status register named $cut$ used to label a given router component as a \textit{cut-vertex} or not, a built-in 
combinational circuit to determine $cut$ values, cf., \textsc{lemma 2} in section~\ref{sec:sda}, and a one-bit status register, assigned at the beginning 
of the \textit{depth-first search}, to indicate whether or not the component is $root$. 
The \textit{Self-Monitoring}'s responsibility is to monitor network activities and to record fault induced topology changes and 
network connectivity. It is also responsible for sending and forwarding potential lost packet messages to neighboring routers for retransmission after self-healing reconfiguration. 

\subsection{\textbf{Self-Reconfiguring unit}}
\label{sec:SRU}
The \textit{Self-Reconfiguring} block shares the \textit{Self-Awareness Module}'s \textit{neighbor list table} with 
\textit{Self-Monitoring} unit and updates it with prohibited routing turns based on the new network connectivity graph constructed 
by the \textit{Self-Monitoring} unit.
Figure~\ref{fig:reconfiguration} depicts the functional view of \textit{Self-Reconfiguring} unit. 

The number of valid neighbors in the table represents the degree of the current node, a \textit{node} can be detected with 
no propagating effects if the number of valid neighbors is 1. Routing algorithms, both \textit{source routing} and 
\textit{node-table routing}, forbid turns to prevent deadlock\cite{glass_turn_1992}\cite{wu2003fault}. 
In the \textbf{\textit{Fashion}} architecture \textit{node-table routing} is adopted because it is more amenable to the 
distributed decision structure in the architecture. The lower half of Figure~\ref{fig:reconfiguration} represents an extension 
of routing table information; it stores the outgoing ports a packet should take to reach the next downstream node. 
If West-to-South turn is faulty, or disabled for deadlock-freedom purposes at the current node, then the table entry is 
set to ``0'' to indicate that it is not allowed with `W' and `S' as horizontal and vertical indexes. The table entry is set 
to ``1'' when the turn is permitted. The comparator circuits are used to test the number of valid neighbors to a given 
node in the network.

In general, node-table based routing may lack scalability. However, the node-table inside the Self-Monitoring unit is designed to store information about permitted (or forbidden) turns at the router. This approach is different from the conventional node-table based routing, where the router generally contains routing information for all the flows that pass through it. Therefore, our node-table size remains constant for the same topology, even with different network sizes.

When a new permanent fault is detected by the \textit{Self-Monitoring unit} and it must be reflected in a modified topology. The \textit{Self-Reconfiguring unit} will be triggered when the newly discovered fault is identified as a \textit{cut-element}. The details of the algorithm are introduced in Section~\ref{sec:SR_alg}. 

\begin{figure}[t]
\begin{center}
\includegraphics[width=0.8\linewidth]{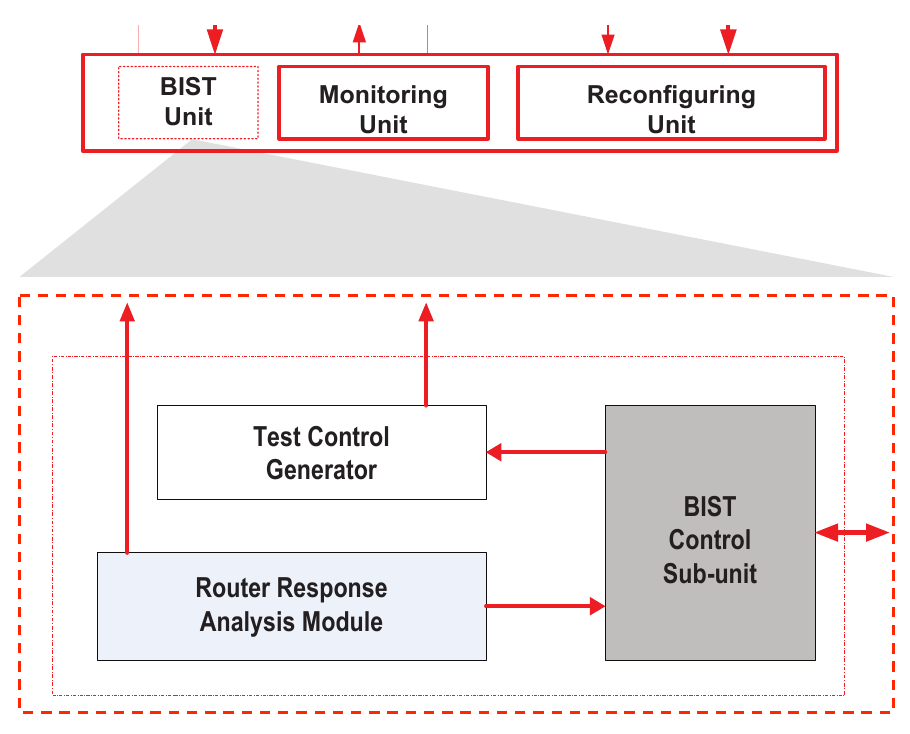}
\caption{\textit{Built-In Self-Test Unit} overview.}
\label{fig:bist}
\end{center}
\vspace{-0.2in}
\end{figure}

\subsection{\textbf{Built-In Self-Test unit}}
The embedded Built-In Self-Test (\textit{BIST}) unit is very standard. It sends test signatures to its neighbors and compares 
received signatures from different directions. If for a given period of time, the \textit{BIST} unit receives no signature response from a neighbor or notices 
that the received acknowledges are different, then it labels the node as disconnected and sets the ``valid'' bit of the corresponding \textit{neighbor list table} entry to ``0''.
Having a \textit{BIST} block located in each router not only decouples the processor element testing from the router\cite{deorio2012reliable}\cite{shamshiri2011end}, but it also allows the unit to be used by \textit{Self-Monitoring} unit to probe links and router internals 
status. Figure~\ref{fig:bist} shows the \textit{BIST} unit in the \textbf{\textit{Fashion}} architecture. 

The BIST unit can also be used in the NoC power-gating domain. A neighbor that is power-gated will be unresponsive and act like a transient fault. The remaining steps of the \textbf{\textit{Fashion}} design remain the same. When the neighbor is turned on again, it will be added back to the connected sub-graph and the reconfiguration algorithm run again. Thus, without loss of generality, we present our analysis assuming unresponsive elements are due to faults.

\section{Fashion Fault Detection and Reconfiguration Algorithm}
\label{sec:algo}
In this section, we present the fault diagnosis and recovery algorithm executed by the \textit{Built-In Self-Test}, \textit{Self-Monitoring}, and \textit{Self-Reconfiguration} units.
The \textit{root} of the graph corresponds to the \textit{System Manager} and is always active. The algorithm starts and terminates at the \textit{root} node. The root node is selected in a manner that maximizes its degree. The \textit{average degree} of a graph G is $d(G) = {\frac {1} {|V|}} \sum_{v \in V} d(v)$ and $\delta(G)\leq d(G)(v)\leq \Delta(G)$, where $\delta(G)$ and $\Delta(G)$ are the \textit{minimum degree} and \textit{maximum degree} of the Graph. This selection method improves the root node's aliveness and connectivity probabilities. 
 
\subsection{Fault Detection and Classification Algorithm}
\label{sec:monitor}
The \textit{Self-Monitoring} unit periodically executes the \textit{BIST} unit. When a new failed component is detected. 
The \textit{Self-Monitoring} needs to verify whether the underlying network is still connected. Tarjan\cite{tarjan1972depth} introduced a central \textit{cut vertex} detection scheme using \textit{depth-first search}. 
However, Tarjan's approach is not able to detect \textit{cut edges} and is not suitable to directly implement in a distributed system such as a NoC. The \textbf{\textit{Fashion}} detection algorithm works in a distributed mode and simultaneously detects both \textit{cut vertices} and \textit{cut-edges} for dynamically changing network topologies. 
The algorithm is executed at network \textit{SAMs} layer. 

\subsection{\textbf{\textit{Self-Monitoring} Algorithm}}
\label{sec:sda}
The \textit{Self-Monitoring} procedure has two phases:
\begin{enumerate}
\item \textbf{Construct the \textit{depth first tree}:}  
During the depth-first traversal the positions and relationships of network nodes and links are recorded;
\item \textbf{Identify the \textit{cut elements} and \textit{connected subgraphs}:}
Nodes and edges are identified and labeled whether they are \textit{cut vertices} and \textit{cut edges} according to 
their location established during the search. 
\end{enumerate}

For a node $i$, $i_{depth}$ is the distance from the node to \textit{root}; $i_{low}$ is the minimal $depth$ value among 
the node, its \textit{non-parent} neighbors and the minimal $low$ value of its \textit{children}:  

\begin{eqnarray*}
\label{eq:low}
i_{low}&=&min(i_{depth}, min(k_{1_{depth}},k_{2_{depth}} ...), \\
      &&{} min(l_{1_{low}},l_{2_{low}} ...)) \quad\quad\quad\quad\quad\quad (1)
\end{eqnarray*}

Where, $k_1,k_2,... \in i_{neighbors} \setminus i_{parent}$, and $l_1,l_2,... \in i_{children}$. 

\textbf{Construction of the \textit{depth first tree}:} 
This procedure is initiated and terminated at both the \textit{root} node. The \textit{counter} records number of \textit{connected components} after the execution. Each node is either \textsc{unvisited} or \textsc{visited} during the process. Already explored nodes are marked as a \textit{child} of previous visited node, in other words, there is one and only one \textit{parent} entry in the \textit{neighbor list table} set to $1$ at a \textsc{visited} node. Thus the \textit{parent} entry state implicitly indicates whether the current node is \textsc{visited} or not.

At the beginning, all the vertices of the network are \textsc{unvisited} and all the variables are initialized (noted as 
$x_{cut}=True$, $x_{root}=False$, $x_{depth}=x_{low}=N$, $counter=0$, $x_{parent}=Null$, $x_{child}=Null$ 
and $x_{neighbors}$ contains the node ids of all its one-hop neighbors, the latter three pieces of information are stored 
in \textit{neighbor list table}, $N$ is the number of total nodes). From the hypervisor or the system manager, 
the operating system launches the construction of \textit{depth first tree}.

The source vertex $i$ is marked as the \textit{root} ($i_{root}=True$), then $i_{depth}$ is also updated 
($i_{depth}=0$) and the $counter$ is set to $1$. Vertex $i$ executes \textit{forward search} to explore 
\textsc{unvisited} nodes as ``deep'' as possible, $i$ sends a forward message to one of its \textsc{unvisited} neighbor 
$j$ in the \textit{neighbor list table} and marks $j$ as $i$'s \textit{child} ($i_{child}=j$) and passes $i_{depth}$ and $counter$ to $j$. At node $j$, 
$i$ is marked as $j$'s \textit{parent} ($j_{parent}=i$) and the update $j_{depth}=i_{depth}+1$ is performed. The $counter$ 
variable is also incremented by one ($counter=counter+1$). 

The exploration of \textit{unvisited} neighbors at $j$ is 
then launched. When $j$ does not have any \textsc{unvisited} neighbors, it executes the \textit{backward search} to discover \textsc{unvisited} neighbors of $j_{parent}$ ($i$). It updates $low$ values of all the nodes along the \textit{backward} paths. Figure~\ref{fig:monitor} shows the ``neighbor-to-neighbor'' communication infrastructure and how recording and updating of all the necessary informations are done during the \textit{depth first tree} searching.

During the \textit{backward search}, $j$ sends a \textit{backward} message to $i$($j_{parent}$). If vertex $i$ is the 
\textit{root} and all of its \textit{children} are already explored, then the \textit{depth-first search} algorithm has walked 
through all the vertices belonging to the \textit{depth first tree}. 
If $i$ is a \textit{non-root} vertex and there are no \textsc{unvisited} nodes inside $i_{neighbors}$, the algorithm moves up to $i$'s parent and $i_{low}$ is updated according to formula (1). 
The $i_{low}$ value is passed to $i_{parent}$ and \textit{depth-first search} at $i_{parent}$ continues until all the vertices are visited. 
The \textit{connected subgraph} initialized from the hypervisor represents the \textit{maximal connected subgraph}, also called the connectivity map of network. The other \textsc{unvisited} nodes are disconnected from the system manager and are labeled as out-of-service, since they cannot be reached or used. The gray components in Figure~\ref{fig:monitor} show the \textit{backward search} logic, including the comparison operation in formula (1) and the setting of the \textit{cut}-bit. In Section~\ref{sec:ex_fashion}, we further demonstrate how the connectivity of the network can be improved by leveraging the bidirectional links and unified virtual channels. 

\begin{figure*}[t]
\begin{center}
\includegraphics[width=0.95\textwidth]{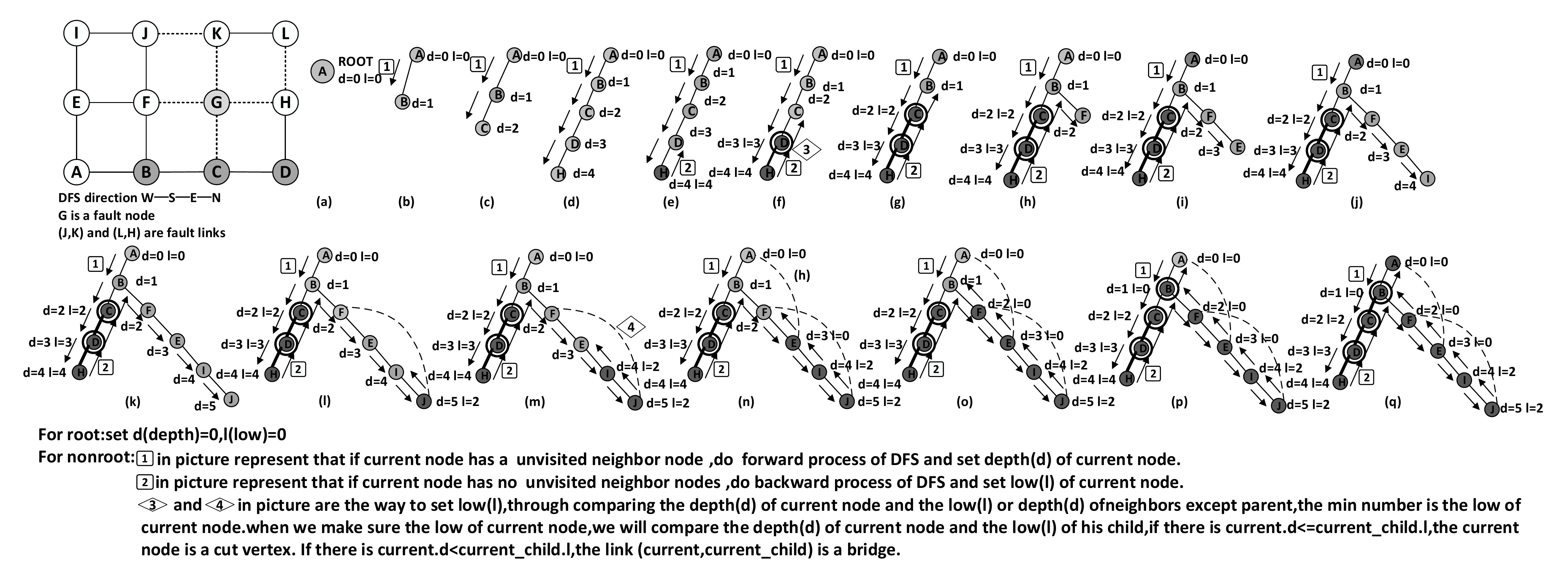}
\caption{Illustrative example of the classification procedure. Node A is the hypervisor that holds the operating system, node B, C, D are \textit{cut vertices}, and edge (B,C), (C,D), (D,H) and (L,K) are \textit{cut edges}. The \textit{cut components} are heavily shaded. The \textit{depth first tree} contains nine vertices. The search started and terminated at \textit{root} A. The other two vertices K and L are disconnected to node A and are labeled as out-of-service.}
\label{fig:example}
\end{center}
\vspace{-0.2in}
\end{figure*}

\textbf{Criterions to identify cut elements}
Assuming $G_{\pi} = (R,L_{\pi})$ is the \textit{depth-first forest} of $G$, then the following two rules are used to 
classify the system nodes as \textit{cut} or \textit{non-cut elements}\cite{cormenintroduction}. 
A \textbf{back edge} is an edge that directly connects a node to its \textit{ancestor}, and there also exists another path from its \textit{ancestor} to the node itself through \textit{tree edges}.  

\begin{itemize}
\item \textbf{\textsc{lemma 1}. Cut vertex detection}: For a vertex $i$, $i_{cut}=True$ if $(i_{root}=True)~\&~$(number of $i_{children}$ $\geq 2)$, or $(i_{root}=False)~\&~$$(\exists j \in i_{children}$, $i_{depth}\leq j_{low})$;
\item \textbf{\textsc{lemma 2}. Cut edge detection}: For an edge $(i,j)$, where $j_{parent}=i$, $(i,j)_{bridge}$=$True$ \textit{if and only if} $((i,j)$ is a \textit{tree edge}$)~\&~(i_{depth} < j_{low})$. 
\end{itemize}
\textbf{Proof of \textsc{lemma 1} (\textit{cut-vertex} detection)}:
The first part of \textsc{lemma 1} is obvious, if $i$ is \textit{root} and it has multiple \textit{children}, it is the \textit{cut-vertex} and vice versa.

Let us consider the case where $i$ is a \textit{non-root} \textit{cut vertex}. There $\exists j\in i_{children}$ and there is no \textit{back edge} from $j$ or any $descendant$ of $j$ to a proper $ancestor$ of $i$. Thus, $i_{depth}$<$j_{depth}$ and $i_{depth}$ is less than the $depth$ value of any neighbor of $j$. If $j_{children} \neq$ {\O} and $\forall m\in j_{children}$, then there is no \textit{back edge} from $m$ to any $ancestor$ of $i$ since $m$ is also a $child$ of $i$. It follows that $i_{depth} \leq m_{low}$. Otherwise, if $j_{children} =$ {\O}, $j$ will be a \textit{leaf node} and $i_{depth} \leq j_{low}$.

The proof of the converse, $i_{depth} \leq j_{low}$ goes as follows: according to the definition of $low$, (a) $i_{depth} \leq j_{depth}$, (b) $i_{depth} \leq n_{depth}$, $\forall n \in j_{neighbors}\setminus i$, and (c) $i_{depth} \leq m_{low}$, $\forall m \in j_{children}$. Statements (b) and (c) guarantee that no \textit{back edge} from $j$ ($j=i_{child}$) or any descendant of $j$ to a proper ancestor of $i$. $\blacksquare$

\textbf{Proof of \textsc{lemma 2} (\textit{cut edge} detection)}:
If edge $(i,j)$ is a \textit{cut edge}, then $i_{depth} = j_{parent_{depth}}$<$j_{depth}$. According to the property of \textit{cut edge}, 
$(i,j)$ does not lie on any cycle, thus there is no edge from $j_{neighbors}$ and $j_{children}$ to $i_{ancestor}$. Therefore, $j_{low} = j_{depth}$>$i_{low}$.

To prove the converse, let us assume that $(i,j)$ is the \textit{tree edge} and $i_{depth}$ < $j_{low}$, then according to formula (1), we have:
(a) $j_{depth}$>$i_{depth}$, because $j_{parent}=i$; (b) $min(m_{depth})$>\\$i_{depth}$, $\forall m\in j_{neighbors}\setminus j_{parent}$, there is no \textit{back edge} from $j_{neighbors}$ to $ancestor$ of $i$; (c) $min(n_{low})$>$i_{depth}$, $\forall n\in i_{children}$, there is also no path connecting $j_{children}$ to $i_{ancestor}$ without passing through node $i$. As a result, $(i,j)$ did not form a cycle and the converse part of the \textsc{lemma 2} is true. $\blacksquare$

\subsection{\textbf{\textit{Self-Reconfiguring} Algorithm}}
\label{sec:SR_alg}
This part of the algorithm is designed to remove deadlock and preserve the connectivity inside the $G^{max}$. Here, we construct an acyclic channel dependency graph (ACDG) that breaks all cycles to avoid in-flight packets becoming trapped in a cyclic pattern, while preserving connectivity by prohibiting turns only at \textit{non-cut elements}. The procedure consists of: 
\begin{enumerate}
\item Identifying the \textit{leaf} nodes set $S_{leaf}$ in the $G^{max}$. Every connected node can independently determine whether or not it is a \textit{leaf} node by checking whether the \textit{child} entry in its \textit{neighbor list table}, see Figure~\ref{fig:reconfiguration}; 
\item Identifying the \textit{non-cut vertices} set $S_{ncut}$ in the $G^{max}$ with minimal degree (for example, the minimal degree is 2 for 2D-mesh/torus network). This can also be achieved individually by counting the number of connected neighbors in the \textit{neighbor list table}. Then, we can forbid turns in the form of $(i,x,j)$ and $(j,x,i)$ at node $x$, $x \in S_{ncut}$ ($x_{cut}=0$) and $i,j \in x_{neighbor}$. Prohibited turns appear in pairs and are recorded in the \textit{node-based routing table}, see Figure~\ref{fig:reconfiguration};
\item Removing $S_{leaf}$ and $S_{ncut}$ from $G^{max}$ by updating the \textit{valid}-bit to $0$ for all their connected neighbors. Since all the removed nodes are \textit{non-cut vertices}, deleting them does not affect the connectivity of the rest of nodes in $G^{max}$; 
\item Constructing a new $G^{max}$ by running another round of spanning tree search. After that, repeat these procedures until the remaining vertices number of $G^{max}$ is equal to 2 ($counter = 2$). Notice that, removing \textit{non-cut elements} does not affect the connectivity of rest nodes in the $G^{max}$. The modification of the topology of $G^{max}$ will however change the \textit{cut vertex} set and degrees of nodes. Therefore, another around of spanning tree search executed by \textit{Self-Monitoring Unit} is needed;
\end{enumerate}
Strictly forbidding turns only at \textit{non-cut elements} position will not destroy the connectivity of the network. 
The proof for the deadlock-freedom property of the proposed ~\textit{Self-Reconfiguring} algorithm is done by reductio ad absurdum:  

It is well known that a routing algorithm is deadlock-free if the nodes can be numbered and messages can only traverse nodes in a strictly decreasing (or increasing) order. So, assuming that the removed nodes in the same iteration have a unique label, and this label is increased at the beginning of next iteration. Every node will have a label when the algorithm is finished. If there is a cycle $C$ in the network, and node $K$ is with the minimum $label(K)$ in $C$. Then, there exists a turn $(U,K,V)$ in cycle $C$, both $label(U)$ and $label(U)$ are greater than $label(k)$. 

However, according to the second procedure $2)$ in the \textit{Self-Reconfiguring Algorithm}, turn $(i,K,j)$ is prohibited, $i,j \in K_{neighbors}$, which brings a contradiction. Therefore, cycle $C$ is non-existent.  
In this way, we guarantee that the self-reconfiguration procedure supplies a deadlock-free and connectivity guaranteed routing solution inside the \textit{maximal connected graph}. This property has also been proved by our experiments.

\subsection{\textbf{Illustrative case of the classification procedure}}
In this illustrative case, we consider a network consisting of $12$ nodes. There are $ids$ associated with each node, see Figure~\ref{fig:example}. 
Fault links $(J,K),(K,G),(G,F),(G,C),(G,H)$ and $(H,l)$ are marked with dashed lines. \textit{Cut elements} and \textit{cut edges} 
are shaded.
The classification routine takes 22 steps to complete. At the beginning, $A$ is selected as the \textit{root} followed by 
a visit to node $B$ and the update $B_{depth}=B_{low}=1$, shown Figure~\ref{fig:example}(b). 
The search continues until it reaches $H$, where there is no \textsc{unvisited} neighbor. A backward search is then initiated with the appropriate updates $H_{low}=H_{depth}$ and $D_{low}=3$ ($D$ is $H_{parent}$) - Figure~\ref{fig:example}(e). In Figure~\ref{fig:example}(g), the backward search reaches node $B$ and discovers node $F$ not yet visited. 

A new forward search starts and the update $F_{depth}=B_{depth+1}$ is made. The forward search reaches node $J$, where a \textit{back edge} connects $J$ and $F$, and $J_{low}$ is set equal to $F_{depth}=2$ - Figure~\ref{fig:example}(k). 
The next step is a backward move to update $I_{low} = J_{low}= 2$ - Figure~\ref{fig:example}(l). At node $E$, the 
algorithm discovers that edge $(A,E)$ connects $E$ to the $root$, and the result is $E_{low} = min(E_{depth}, A_{depth}, 
I_{low}) = min(3,0,2) = 0$ according to formula (1), shown in Figure~\ref{fig:example}(m). 
The \textit{depth first tree} rooted at $A$ has 9 vertices - Figure~\ref{fig:example}(p), which forms the \textit{maximal connected subgraph}. 
There are still \textsc{unvisited} (disconnected) nodes $K$ and $L$ in the network, which are labeled as out-of-service.

\subsection{\textbf{Algorithm complexity analysis}}
In the algorithm, each edge in the network is traversed at most twice in the building of the \textit{depth first tree}. The 
upper bound of the algorithm is therefore $2L$ steps, where $L$ is the number of edges in the network. 
The part of the algorithm that executes on the \textit{Self-Monitoring} unit needs $O(|L|)$ time to finish. 
In the \textit{Self-Reconfiguring} unit, every node is checked to determine \textit{leaf} nodes. Prohibiting turns may prompt re-runs of this 
part of the algorithm. Therefore, the \textbf{\textit{Fashion}} algorithm has worst-case computational cost of 
$O(|R||L|)$. For a $N \times N$ 2D-mesh network, $|L|=4N(N-1)=4N^2-2$, $|R|=N^2$, and the computational 
complexity is $O(4|R^2|)$. In the conducted experiments, 
it took an average of 706.49 and 1384.45 clock cycles to finish the 
scheme on networks of 64 and 256 nodes.

\subsection{\textbf{System deployment senario}}
Researchers have realized that addressing the challenges of permanent errors in NoC requires combining both hardware and software efforts. While hardware is required to recover from permanent defects and self-adjust to continually guarantee correct operation, the software is responsible for eliminating the fault-induced performance impacts\cite{borkar2005challenges}\cite{sylvester2006elastic}\cite{deorio2011drain}\cite{radu:2009}.

At runtime some information may be exchanged between the operating system (OS) and the \textit{Self-Awareness Module} (\textit{SAM}). However, the proposed algorithm is executed entirely by the \textit{SAM} unit. The OS can trigger and interrupt the execution of the algorithm. For example, the OS can periodically activate the BIST or \textit{Self-monitoring units} to obtain connectivity information before assigning new applications (or threads) to available processing elements.
The monitoring, reconfiguration and built-in testing sub-units are programmed before the start of the application execution alongside the routing tables and virtual channel allocation algorithms. 

During the application execution, the \textit{SAM} unit is activated periodically, meanwhile, architecture states and other parameters are periodically in memory. When a fault is detected, packets and flits are stalled at their destinations. Using functional links, potential lost or corrupted packet messages are sent back to neighbors and propagated from there depending on the criticality of the fault. The \textit{SAM} unit then goes through its local self-healing/reconfiguration process. When it is done, credits are updated and normal application traffic can resume going through the router. The restoring mechanism rolls back to a previous error-free execution point, and the system moves forward to the next state.

In general, the software stack is oblivious to the online detection and reconfiguration. The system software may be involved if substantial 
part of the chip becomes faulty, necessitating a full system pause and global reconfiguration e.g., a time-out mechanism for dropping on-the-flight packets and issuing retransmission\cite{murali2005analysis}. 

\section{Extended-Fashion Router}
\label{sec:ex_fashion}

\begin{figure}[t]
\begin{center}
\includegraphics[width=0.4\textwidth]{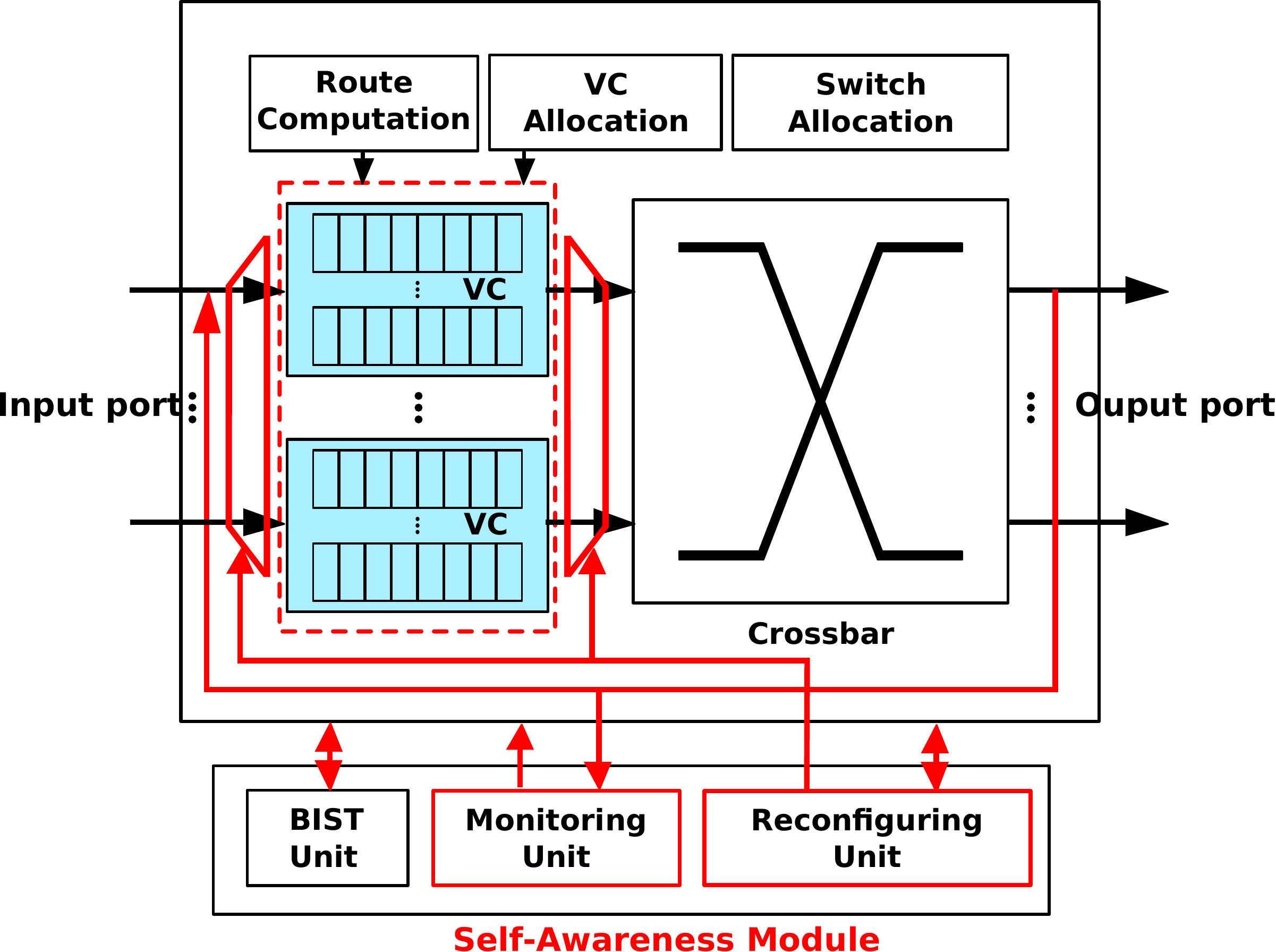}
\caption{Extended-\textbf{\textit{Fashion (Ex-Fashion)}} router.}
\label{fig:exfashion_router}
\end{center}
\vspace{-0.2in}
\end{figure}

To give the network flexibility and adaptation range, two key ingredients are added to the \textbf{\textit{Fashion}} router design, namely, 
coarse-grained reconfigurable bidirectional physical links and unified virtual channel structure, as shown in Figure~\ref{fig:exfashion_router}. 

Cho {\it et al.} \cite{Cho:2009} introduced a bandwidth-adaptive network where the link bisection bandwidth can adapt 
to changing network conditions using local state information. Similar work using bidirectional links is proposed in\cite{Lan2011A}. In these schemes, unidirectional links between network node pairs are merged into a set of bidirectional links. Each new bidirectional link can be configured to deliver packets in either direction. The links can be driven from two different sources, with 
local arbitration logic and tristate buffers ensuring that both do not simultaneously drive the same wire. Tsai {\it et al.}
 \cite{Tsai2012} successfully used this approach to handle static and dynamic channel failures on the data-link layer. 
The~\textbf{\textit{Ex-Fashion}} architecture extends the concept by using the bidirectional nature of links to sustain higher connectivity and rebalance 
link bandwidths around faults in a coarse-grained fashion. The robustness of this approach is drawn from the fact that component failures affecting one  
datapath-segment account for 96\% of faults and can be entirely masked by rerouting around a single link\cite{parikh2013udirec}. 

Virtual channels are another point of vulnerability that is highly susceptible to faults\cite{Dongkook}. 
Instead of having a set of virtual channels strictly associated to a given port as seen in the conventional router, 
\textbf{\textit{Ex-Fashion}} has a pool of virtual channels that can be shared among the ports. It follows the unified virtual 
channel structure approach of the ViChaR\cite{Nicopoulos:2006} router architecture. In this framework, virtual channels 
are not statically partitioned and fixed to input ports, they are rather communal resources dynamically managed by 
the \textit{SAM} unit. This approach allows for faulty buffer to not impact any particular port or render a port unusable. 
In ViChaR, the authors showed that the area and power overheads are negligible and network latency 
can be decreased by 25\% on average using a unified buffer structure scheme. 

In \textbf{\textit{Ex-Fashion}}, the overheads are even lower, because the virtual channel depth is kept constant removing some of 
the complexity associated with the ViChaR architecture. 
Although the use of bidirectional link and unified virtual channel structure are not new, their combination and the 
design of distributed intelligence, i.e., \textit{Self-Awareness Module (SAM)} to dynamically manage them, 
represents one of major contributions of this work. 
The implementation of bidirectional link could improve the connectivity of the network, therefore achieving a better network connectivity map during the self-monitoring phase. 

For example in the network shown in Figure~\ref{fig:example}, if the link from node $H$ to $L$ is broken but the link from $L$ to $H$ is still functional, using the time-division-multiplexed property of the bidirectional link in the \textbf{\textit{Ex-Fashion}} router architecture, we perform direct packet routing between nodes $H$ and $L$. And the \textit{maximal connected graph} would contain two more nodes than \textbf{\textit{Fashion}} architecture. 

Furthermore, a single broken input buffer at a port would not affect the correct functionality of the input port using the unified virtual channel structure, which will further improve the fault-tolerance of the architecture. The possibility of disconnected node is significantly reduced in the new extended router architecture. The \textit{Self-Reconfiguring} module in the \textbf{\textit{Ex-Fashion}} reconfigures the physical links and recomposes the unified virtual channel structure to match the new link directions. 

\section{Evaluation}
\label{sec:eval}

\begin{table*}[t]
\centering
\begin{center}
\begin{tabular}{|c|c|c|c|c|c|c|c|c|} \hline\hline
{Fault} &\multicolumn{4}{c|}{8x8 2D mesh}&\multicolumn{4}{c|}{16x16 2D mesh}\\
\cline{2-9}
{} &\multicolumn{2}{c|}{\textbf{\textit{Fashion}}}&\multicolumn{2}{c|}{\textbf{\textit{Ex-Fashion}}} 
&\multicolumn{2}{c|}{\textbf{\textit{Fashion}}}&\multicolumn{2}{c|}{\textbf{\textit{Ex-Fashion}}} \\
\cline{2-9}
{num}&Avg.\textit{cut-e}&Fully Connected&Avg.\textit{cut-e}&Fully Connected&Avg.\textit{cut-e}&Fully Connected&Avg.\textit{cut-e}&Fully Connected\\ \hline
{10}&{1.064 }&{99.03\%}&{0.057}&{99.99\%}&{0.015}&{  100\%}&{0.014}&{100\%}\\ \hline
{20}&{3.014 }&{95.13\%}&{0.142}&{99.99\%}&{0.191}&{99.83\%}&{0.035}&{100\%}\\ \hline
{30}&{5.996 }&{85.67\%}&{0.247}&{99.94\%}&{0.693}&{99.11\%}&{0.041}&{100\%}\\ \hline
{40}&{10.075}&{68.39\%}&{0.386}&{99.82\%}&{1.714}&{96.83\%}&{0.060}&{100\%}\\ \hline 
{50}&{15.517}&{47.46\%}&{0.554}&{99.70\%}&{3.303}&{91.69\%}&{0.079}&{100\%}\\ \hline
{60}&{22.608}&{26.34\%}&{0.781}&{99.40\%}&{5.553}&{82.05\%}&{0.096}&{99.96\%}\\ \hline \hline
\end{tabular}
\end{center}
\caption{Average number of \textit{cut elements (cut-e)} and the percentage of fully connected nodes in 8x8 and 16x16 2D-Mesh networks.}
\label{tab:statistic}
\vspace{-0.2in}
\end{table*}

\begin{figure*}[t]
\begin{center}
\subfloat[Throughput results for \textsc{uniform-random}.]{
\includegraphics[width=0.33\textwidth,height=1.5in]{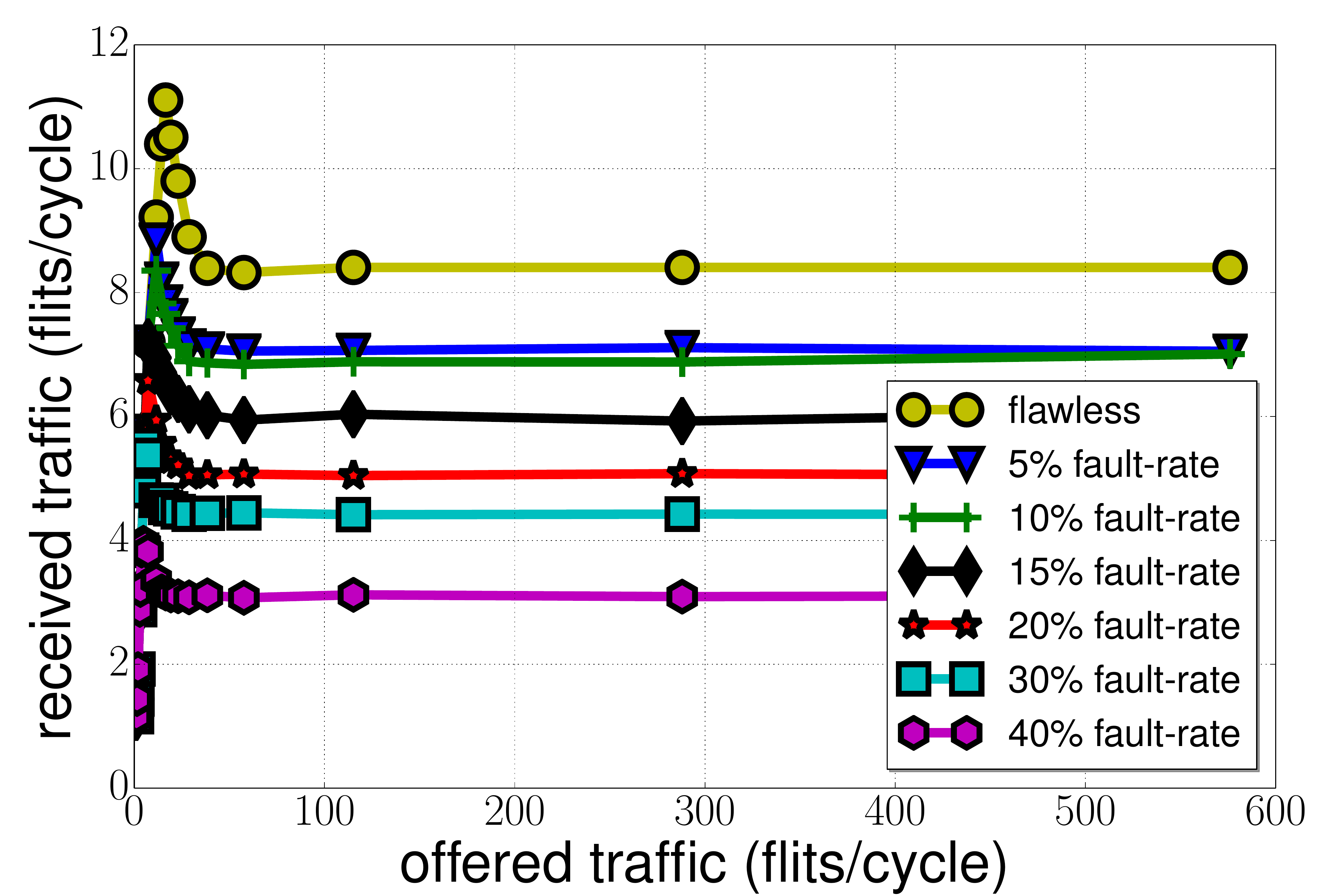}
\label{fig:throughput}}
\subfloat[Throughput results for \textsc{transpose}.]{
\includegraphics[width=0.33\textwidth,height=1.5in]{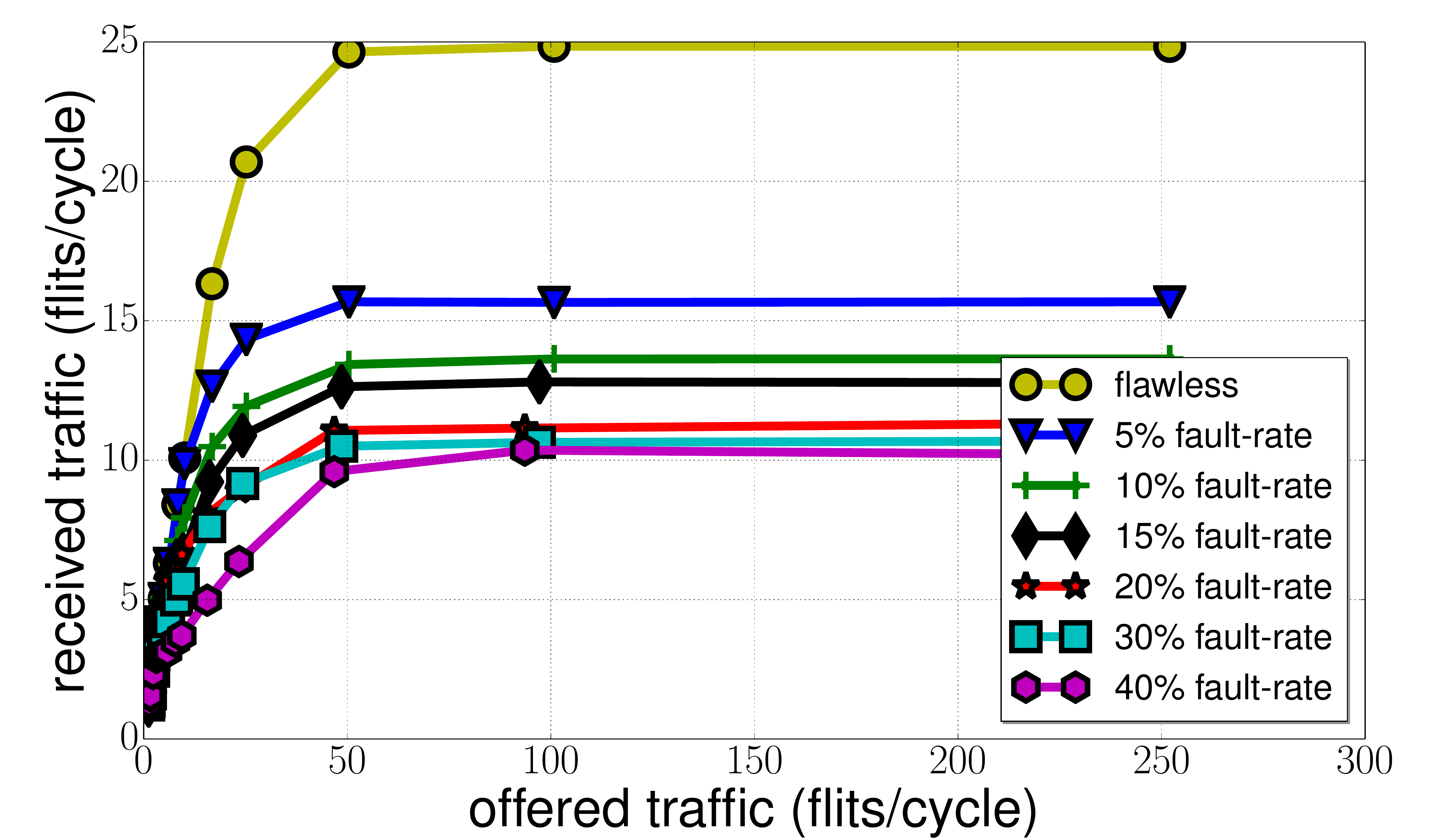}
\label{fig:tr_throughput}}
\subfloat[Throughput results for \textsc{bit complement}.]{
\includegraphics[width=0.33\textwidth,height=1.5in]{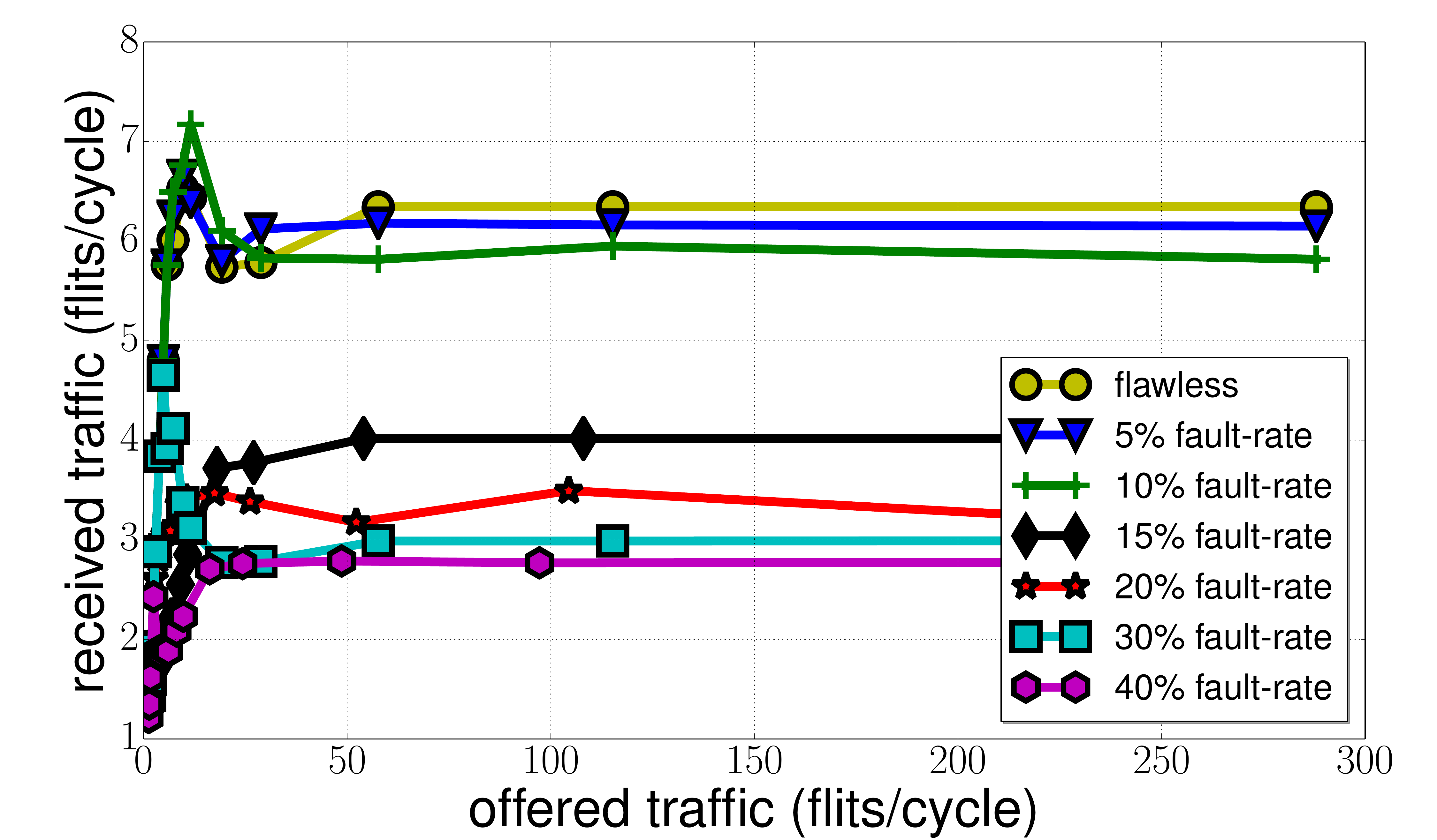}
\label{fig:bc_throughput}}\\
\subfloat[Latency results for \textsc{uniform-random}.]{
\includegraphics[width=0.33\textwidth,height=1.5in]{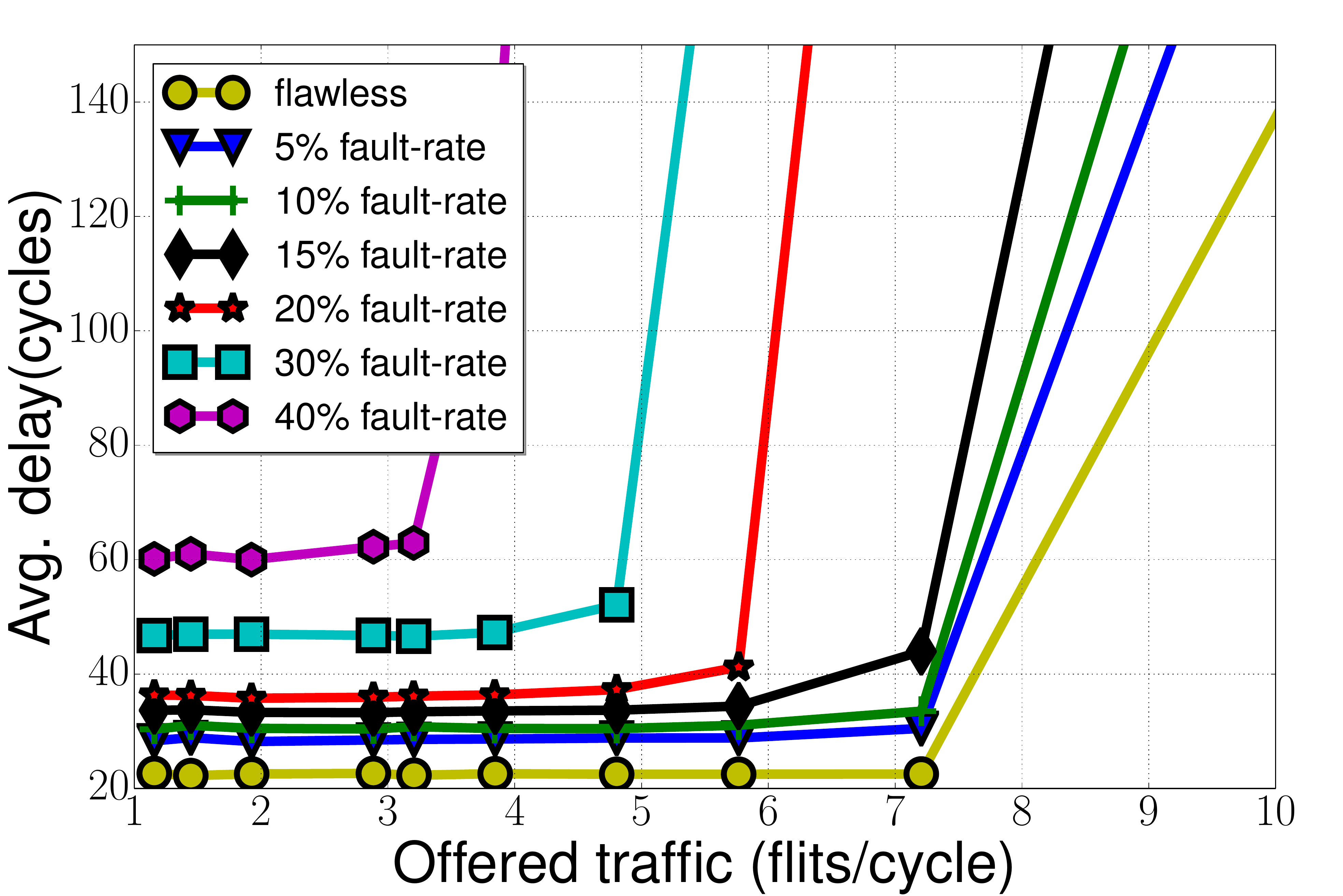}
\label{fig:latency}}
\subfloat[Latency results for \textsc{transpose}.]{
\includegraphics[width=0.33\textwidth,height=1.5in]{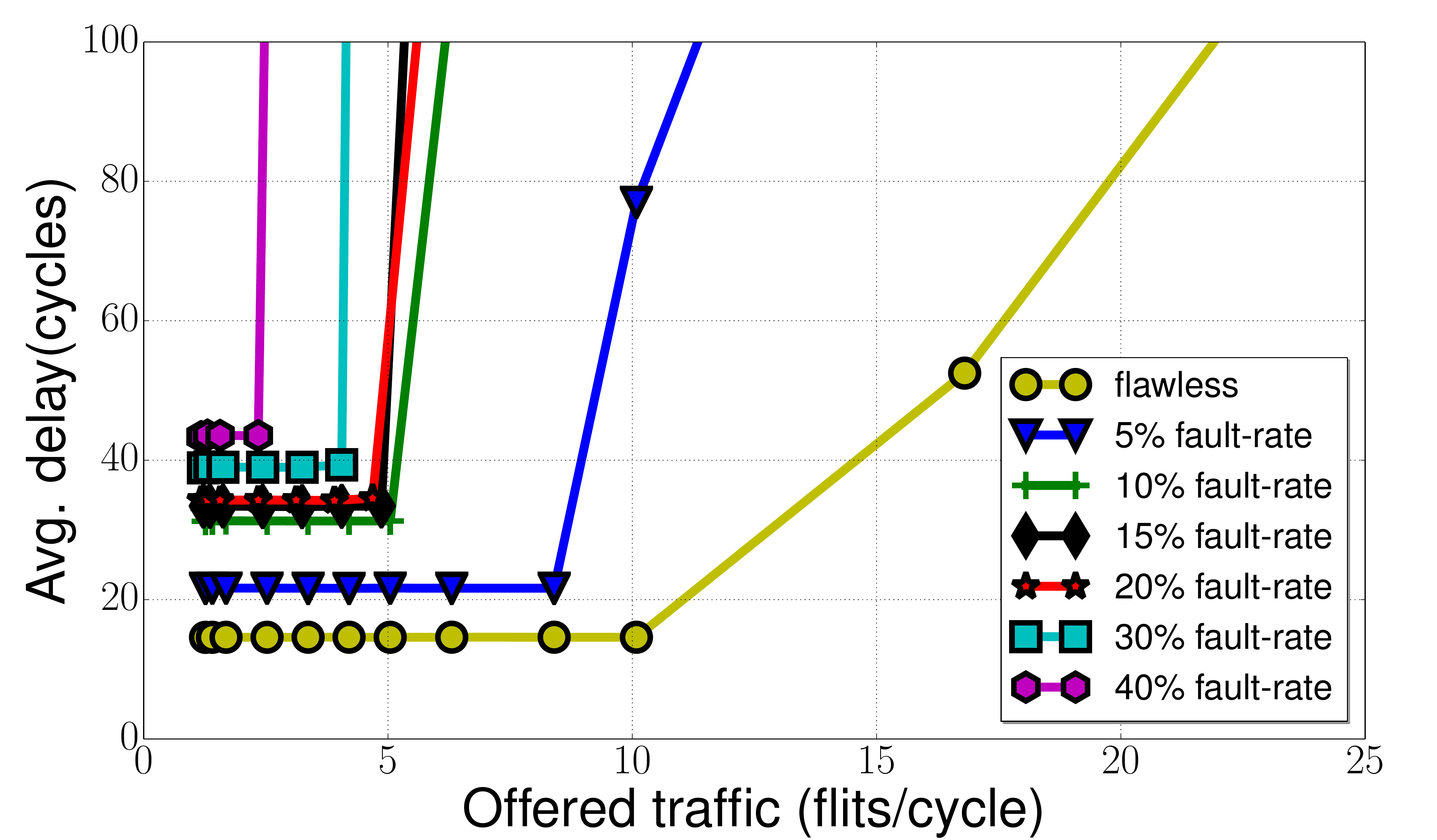}
\label{fig:tr_latency}}
\subfloat[Throughput results for \textsc{uniform-random}.]{
\includegraphics[width=0.33\textwidth,height=1.5in]{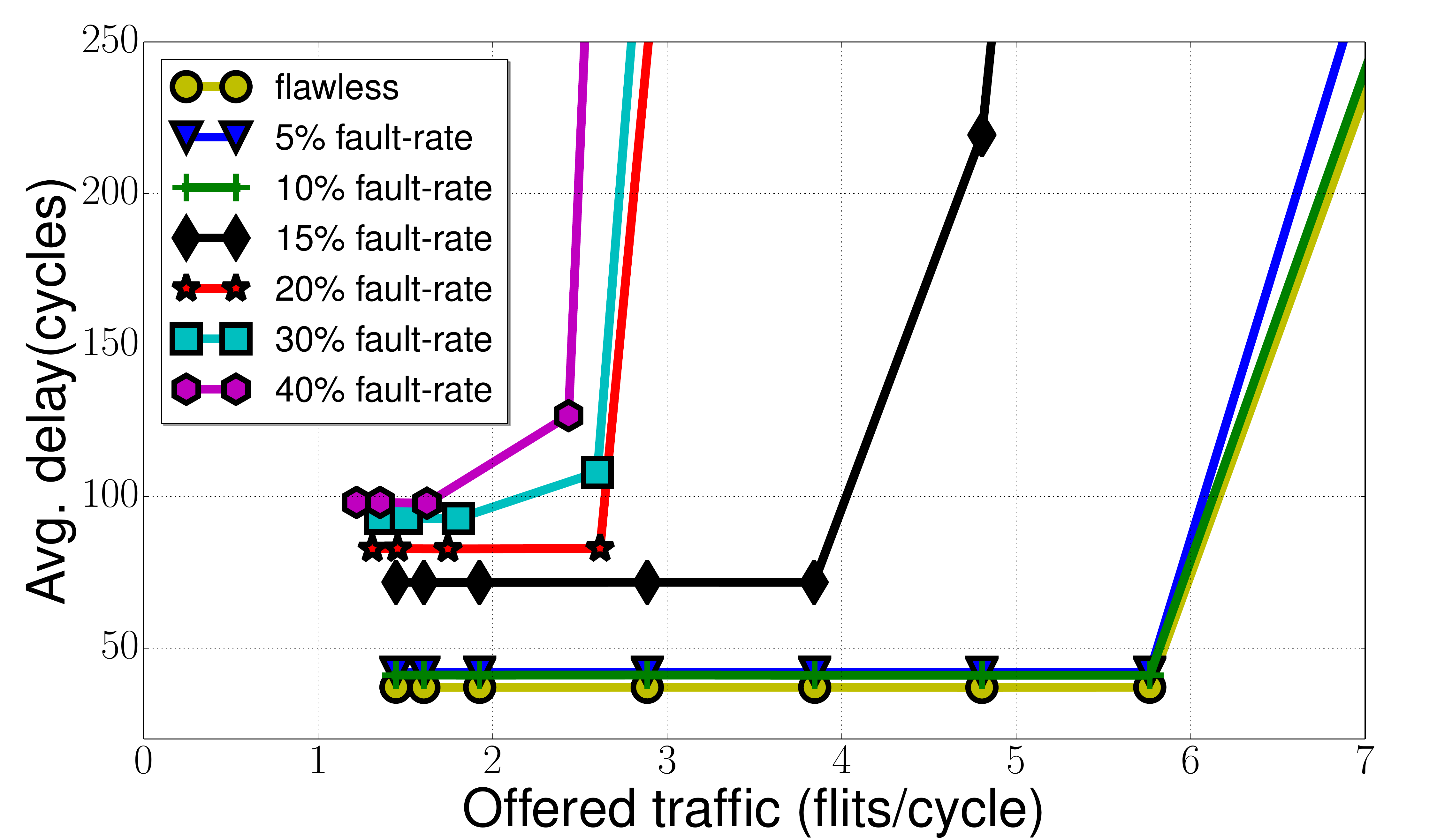}
\label{fig:bc_latency}}
\caption{Synthetic benchmark results for 8x8 2D-Meshes, there are 4VCs for each port and each VC contains 8 flits, packet size is 8 flits.}
\label{fig:th_lat}
\end{center}
\vspace{-0.2in}
\end{figure*}

In this section, we provide extensive performance evaluations of the \textbf{\textit{Fashion}} and \textbf{\textit{Ex-Fashion}} architectures.
\subsection{Hardware Complexity Evaluation}
To accurately evaluate the hardware complexity of the \textbf{\textit{Fashion}} architecture, we 
use the Synopsys Design Complier with TSMC 65nm standard cell library to estimate the architecture area 
overhead. The amounts of hardware required to implement the \textit{self-monitoring} unit and the \textit{self-configuring} 
unit are $2209.32 \mu m^2$ and $3076.2\mu m^2$, respectively. The area is $228620.88\mu m^2$ for a 
64-bit router of an 8x8 2D-mesh, with four virtual channels at each port, and each virtual channel is eight flits depth. 
Thus, the area overhead of \textit{self-monitoring} and \textit{self-configuring} units are 0.966\% and 1.345\%, respectively. 
As for a 16x16 256 2D-mesh, the area percentages are 1.139\% and 1.521\%, respectively. Total area overheads 
become 2.311\% and 2.659\% for \textbf{\textit{Ex-Fashion}} router on 8x8 and 16x16 2D-meshes when the unified virtual channel structure and tristate controller added to the physical links are factored in to the logic cost. 

\subsection{Simulation Details}
We used \textsc{hornet}, a cycle-level many-core simulator\cite{ren2012hornet} for our simulations.
We implemented 8x8 2D-meshes with different numbers of faults. Unpredictable faults may occur at any place in the network, thus we assumed a uniform-random distribution of faults over silicon area. 
As previously mentioned, 96\% of the faults can be masked as broken links\cite{parikh2013udirec} and others are diagnosed as fully broken. For the experiments, the ratio of fault links and fault routers is around 24:1, as described in Section~\ref{sec:connectivity}. 100,000 simulations are performed with various fault numbers and distribution for synthetic benchmarks to explore as many different fault combinations as possible. Applications are reconfigured to run on the \textit{maximal connected subgraph} if the network becomes disjointed due to faults. All experiments have 200,000 warm up cycles and a total of 1,200,000 analyzed cycles.

\subsection{Connectivity Analysis}
Table~\ref{tab:statistic} yields important insight into the large scale system and proposed ~\textbf{\textit{Fashion}}. 
There is a high probability that the network will be split into disjointed subgraphs when the number of faults increase, and the system size makes acquiring a global defect map impractical as the number of nodes increases. An on-line and distributed light-weighted fault-recovery mechanism like \textbf{\textit{Fashion}} and \textbf{\textit{Ex-Fashion}} that can potentially maximize network connectivity is a vital necessity, especially for future massively parallel many-core systems. 

\subsection{Performance Analysis}
In this section, we measure the performance of the \textbf{\textit{Fashion}} architecture under different fault-rates using both synthetic benchmarks and real application traces. 

Figure~\ref{fig:throughput} and~\ref{fig:latency} show the throughput and average latency results of different fault-rates under \textsc{uniform-random} traffic pattern, respectively. Saturation throughput quickly collapses when the fraction of faults increase over 10\% and latency grows rapidly beyond 20\% fault-rate. This is due to the reduced availability of communication resources and congestion caused by faults. 

Figure~\ref{fig:tr_throughput} and~\ref{fig:tr_latency} display the throughput and average latency results of different fault-rates under \textsc{transpose} traffic pattern, respectively. The system throughput drops significantly even at 5\% fault rate. The number of received 
flits per cycle went from $25$ to around $15$. The fact that further system component failures$-$from 5\% to 10\%$-$seems to have a less drastic 
effect on the throughput or latency, highlights the importance of having even limited hardware self-healing capabilities.

Figure~\ref{fig:bc_throughput} and~\ref{fig:bc_latency} show the throughput and average latency results of different fault-rates under \textsc{bit complement} traffic pattern, respectively. The results are in line with the other two benchmark data. Here the most interesting 
aspect is the small degradation effect seen on the throughput and latency results when increasing faults from 20\% to 40\%. 

\begin{figure*}[t]
\begin{center}
\subfloat[Number of faults is 5]{
\includegraphics[width=0.33\textwidth]{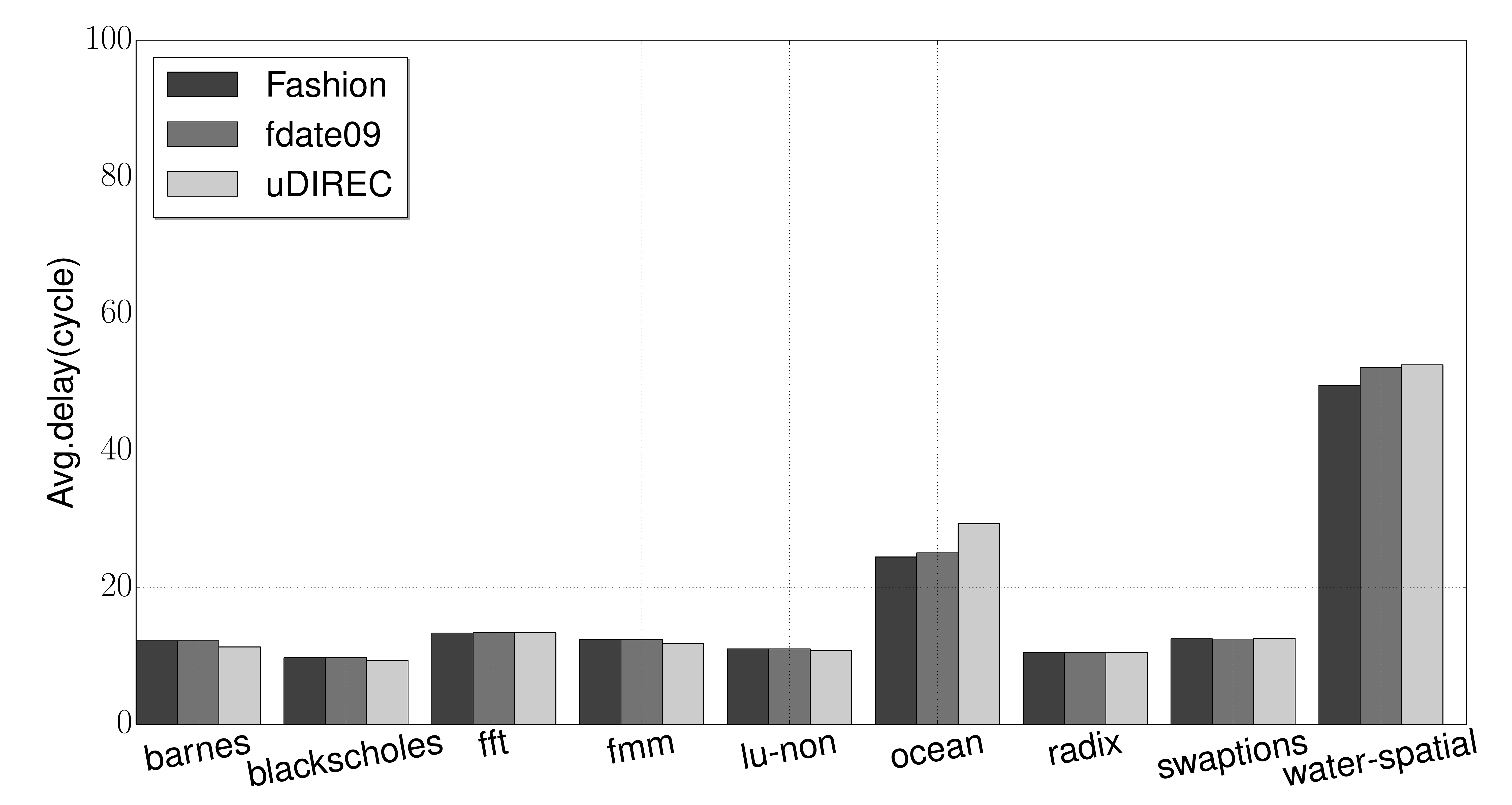}
\label{fig:splash_f5}}
\subfloat[Number of faults is 10]{
\includegraphics[width=0.33\textwidth]{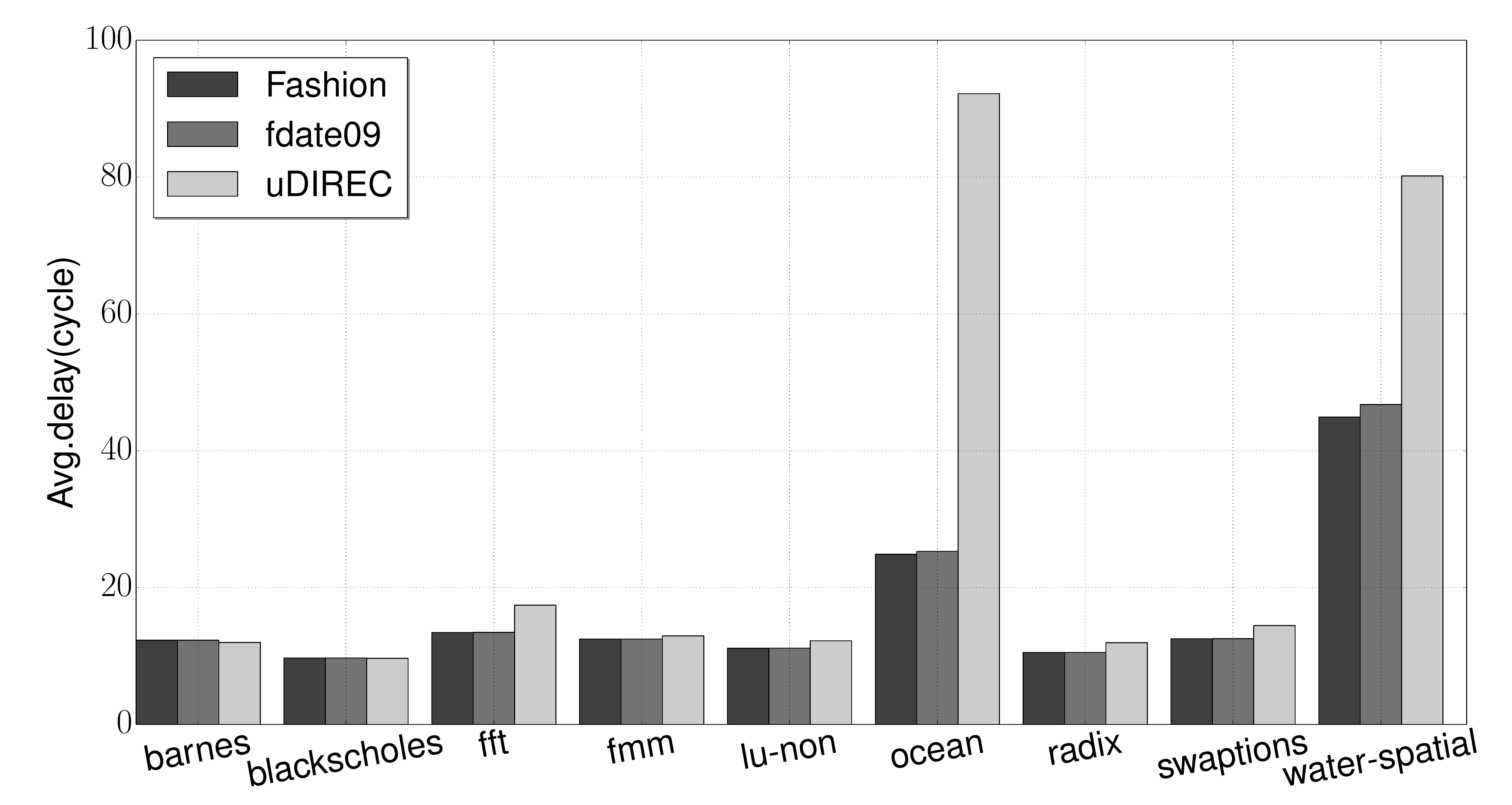}
\label{fig:splash_f10}}
\subfloat[Number of faults is 20]{
\includegraphics[width=0.33\textwidth]{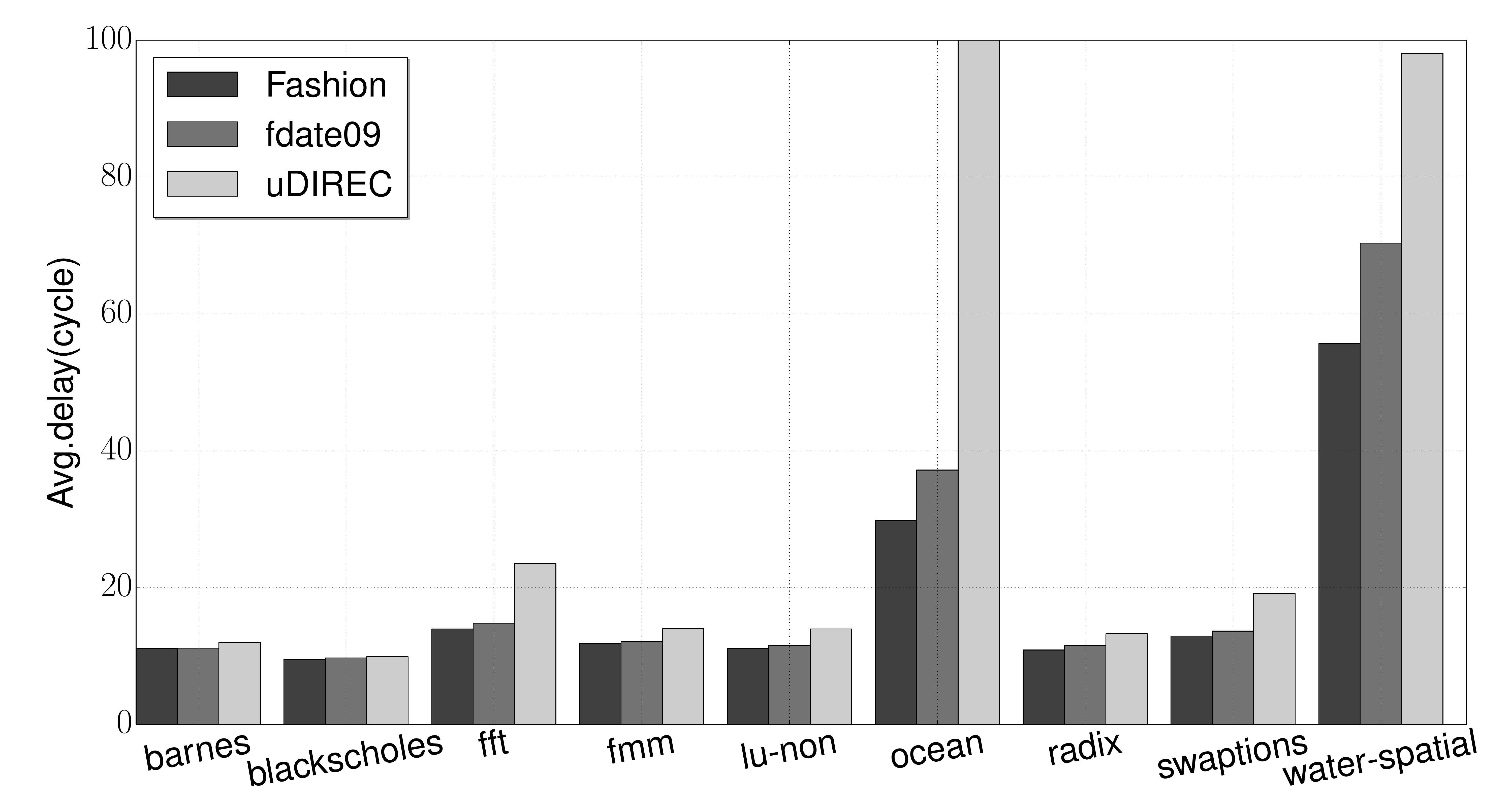}
\label{fig:splash_f20}}
\caption{Average flit latency of selected \textsc{splash-2} benchmarks, the results for the remaining traffic exhibited the same feature and we omit them here for brevity. and number of VCs per port is 4 with 8 flits per VC.}
\label{fig:flash}
\vspace{-0.1in}
\end{center}
\end{figure*}

\begin{figure*}[htb]
\begin{center}
\subfloat[Number of faults is 5]{
\includegraphics[width=0.35\textwidth]{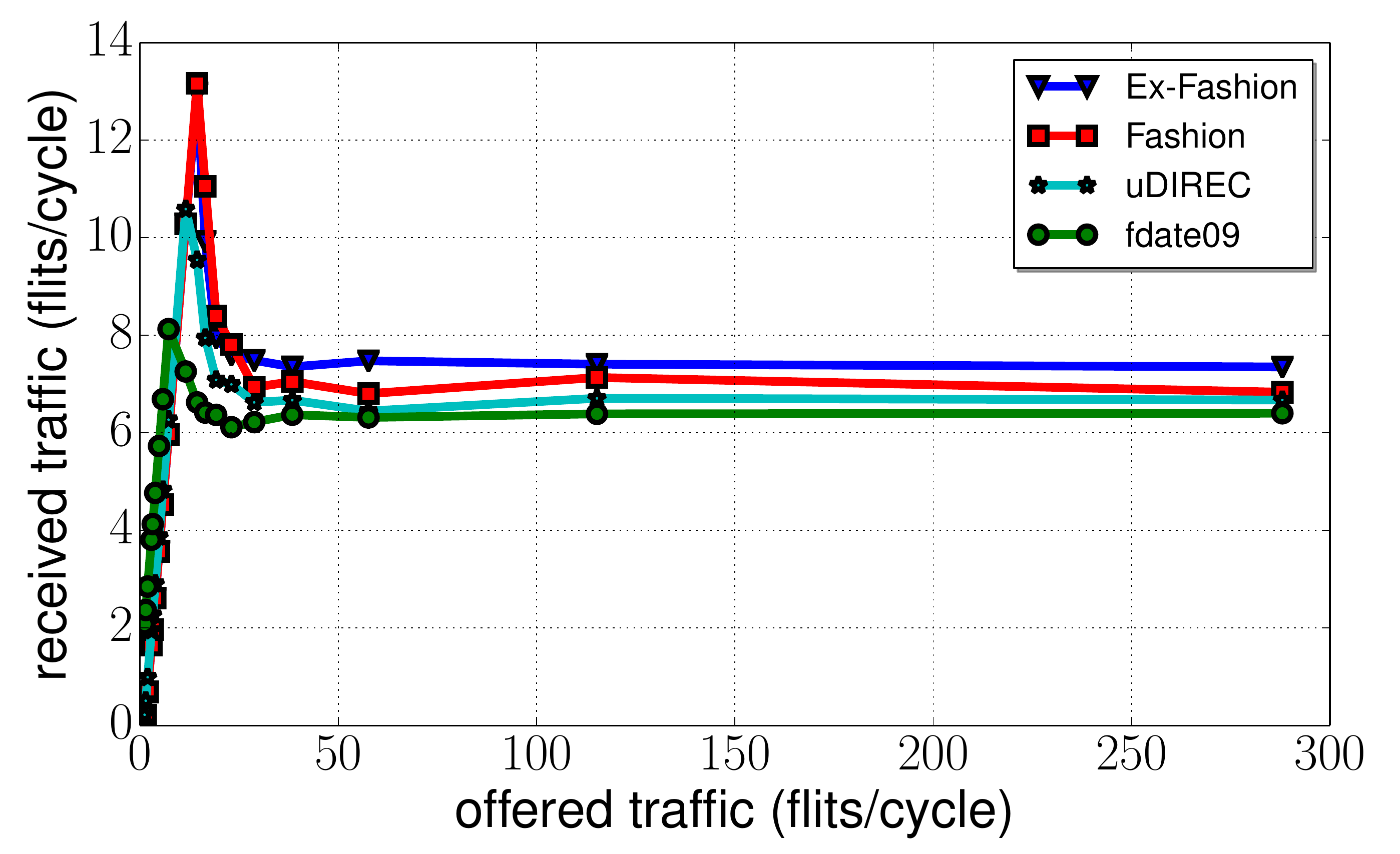}
\label{fig:throughput_f5}}
\subfloat[Number of faults is 15]{
\includegraphics[width=0.35\textwidth]{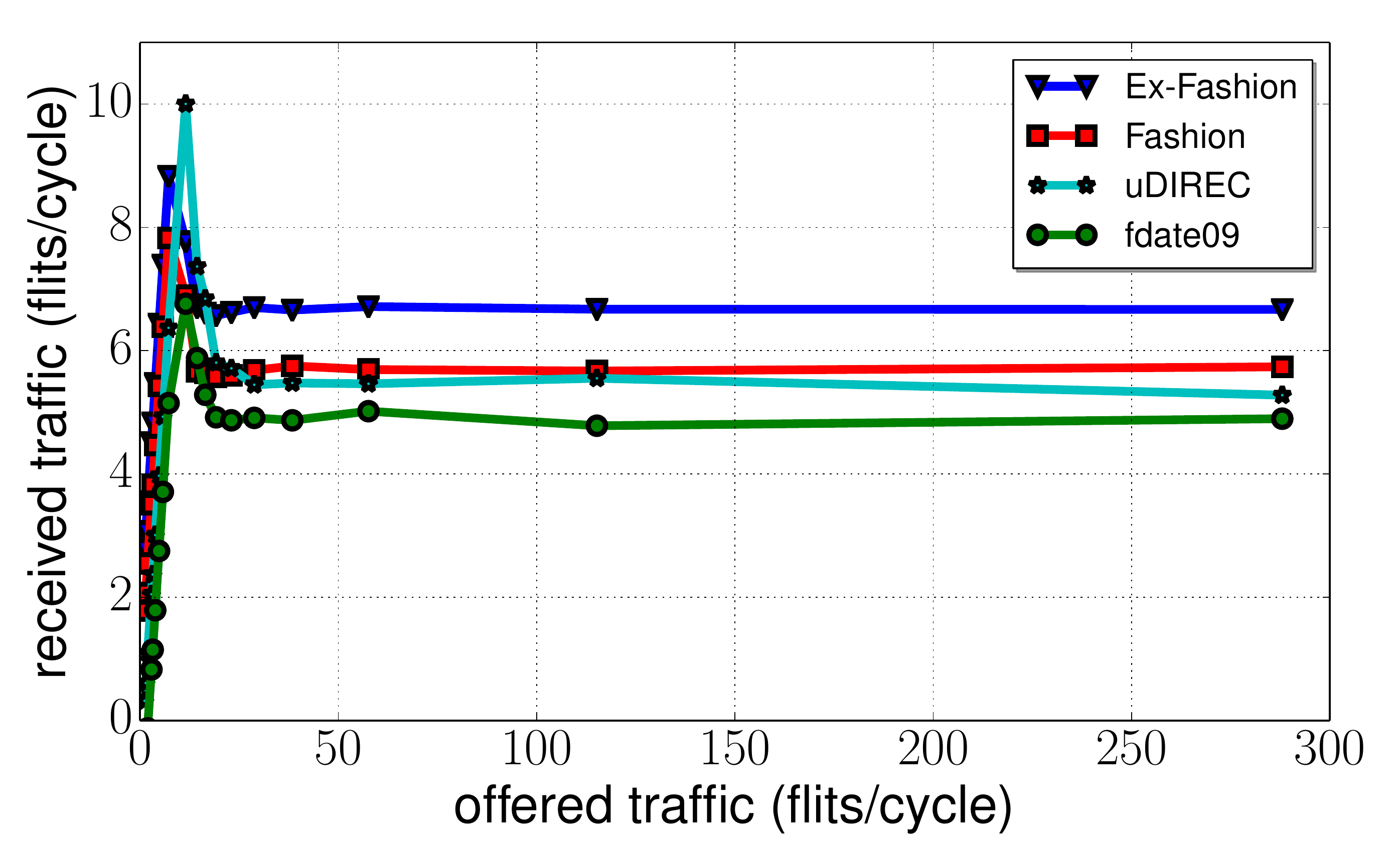}
\label{fig:throughput_f15}}\\
\subfloat[Number of faults is 5]{
\includegraphics[width=0.35\textwidth]{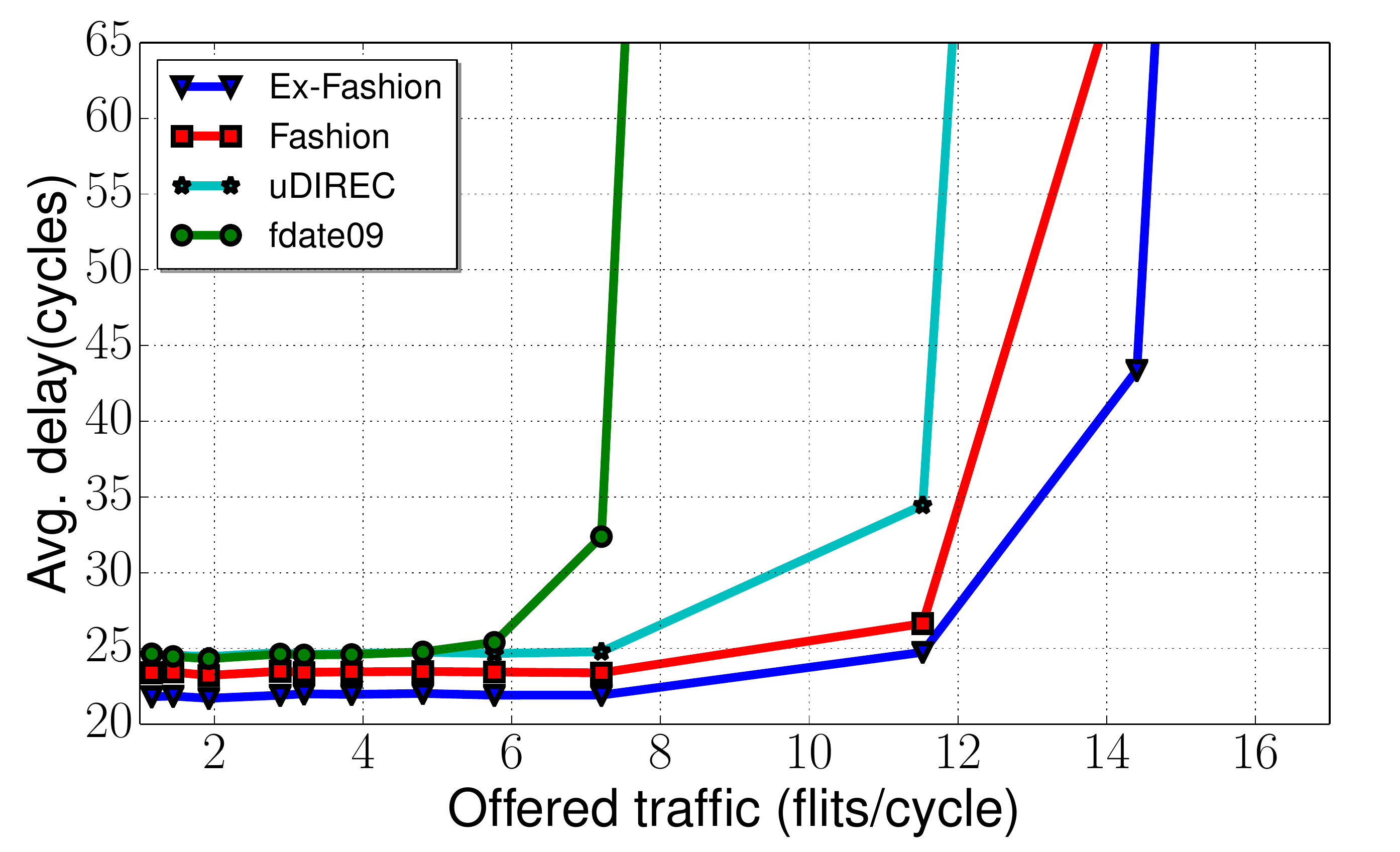}
\label{fig:latency_f5}}
\subfloat[Number of faults is 15]{
\includegraphics[width=0.35\textwidth]{Lat_comp_f5.pdf}
\label{fig:latency_f15}}
\caption{Performance results of 5 and 15 faults for 8$\times$8 2D-Meshes under \textsc{uniform-random} traffic pattern.}
\label{fig:comp}
\vspace{-0.2in}
\end{center}
\end{figure*}

Beyond synthetic traffic, we use traces from the \textsc{splash-2} benchmarks to evaluate the performance of the 
\textbf{\textit{Fashion}} router architecture on 8$\times$8 2D-meshes. The traces are generated from the distributed x86 multicore simulator 
Graphite\cite{miller2010graphite} with 64 application threads. Results for light traffic loads, like \textsc{blacksholes}, are shown in Figure~\ref{fig:splash_f5}-\ref{fig:splash_f20}. 
The average latency remains fairly constant across these benchmarks. For heavy traffic loads like \textsc{ocean\_non\_contiguous} and \textsc{water-spatial}, increasing the number of faults increases the average latency.

\subsection{Comparative Study} 
\subsubsection{\textbf{Baselines}}
For the comparative study, we consider \textit{F(date09)} by Fick {\it et al.}\cite{fick2009highly} and uDIREC\cite{parikh2013udirec} as our baselines. 
\textit{F(date09)} applies flag transmission and routing entry update mechanism, while uDIREC uses a ``supervisor'' to maintain the topology information and according to which, it makes fault tolerance routing decisions.
uDIREC architecture is shown to be more resource and performance efficient 
than previous works \textsc{vicis}\cite{fick2009vicis}, \textsc{immunet}\cite{puente2004immunet} and \textsc{ariadne}\cite{aisopos2011ariadne}. Therefore, we omit them in the following figures for brevity.

\subsubsection{\textbf{Time efficiency}}
uDIREC was proposed to eliminate resource overhead using a scoreboard that keeps the topology 
information at a supervisor node. 
However, as reported in\cite{parikh2013udirec}, network connectivity and performance of uDIREC 
are sensitive to the way \textit{breadth-first trees} are grown. It may take hundreds of milliseconds to finish 
each reconfiguration ($\sim$170ms). It also needs to run multiple iterations to find the optimized results to meet fault-tolerant requirement, which, in turn, 
leads to tens of seconds for a medium sized network (16$\times$16 2D-mesh) and constrains it to a centralized implementation\cite{parikh2013udirec}.

As analyzed in Section~\ref{sec:algo}, the computation complexity of the \textbf{\textit{Fashion}} router is $O(|R||L|)$. 
The results show an average of 1384.46 clock-cycles to complete a $16 \times 16$ 256 2D-mesh 
network, which takes negligible execution time compared with uDIREC. This is 6 to 7 orders of magnitude increase in time efficiency when running at 1Ghz. 

\subsubsection{\textbf{Performance and resource utilization}}
The \textbf{\textit{Fashion}} design is not only more time efficient than the ``off-line'' uDIREC, but it also improves the system performance. 
In most cases, \textbf{\textit{Fashion}} offers lower latency compared to \textit{F(date09)} and uDRIEC. Its latency increases are more gradual than other techniques, especially for heavy traffic loads applications like \textsc{ocean} and \textsc{water-spatial}, Figure~\ref{fig:splash_f5}-\ref{fig:splash_f20}.
The \textsc{SPLASH} benchmarks have relatively low-traffic loads, and do not stress the network performance as much as synthetic benchmarks. Fig~\ref{fig:throughput_f5}-\ref{fig:latency_f15} show the throughput and average flit latency results for 8$\times$8 2D-Meshes using different fault-tolerant mechanisms under \textsc{uniform-random} traffic. The \textit{\textbf{Fashion}} architecture shows a 2.40\% gain in throughput and 4.80\% decreases in average packet latency when compared with uDIREC with 5 faults. The \textit{\textbf{Ex-Fashion}} architecture achieves 11.09\% more throughput and 6.06\% less packet latency under the same experimental condition, Figure~\ref{fig:throughput_f5} and Figure~\ref{fig:latency_f5}. When the number of faults increase to 15, the \textit{\textbf{Ex-Fashion}} design has a throughput of 6.67 filts/cycle compared to the 5.74 filts/cycle and 5.28 filts/cycle for the \textit{\textbf{Fashion}} and uDIREC designs, corresponding to 16.2\% and 26.32\% performance improvement, respectively, over these designs, Figure~\ref{fig:throughput_f15}. In terms of latency, the \textit{\textbf{Ex-Fashion}} and \textit{\textbf{Fashion}} architectures show average packet latency reductions of 18.66\% and 14.67\% over the uDIREC design, Figure~\ref{fig:latency_f15}.    

\begin{figure}[t]
\begin{center}
\includegraphics[width=0.4\textwidth]{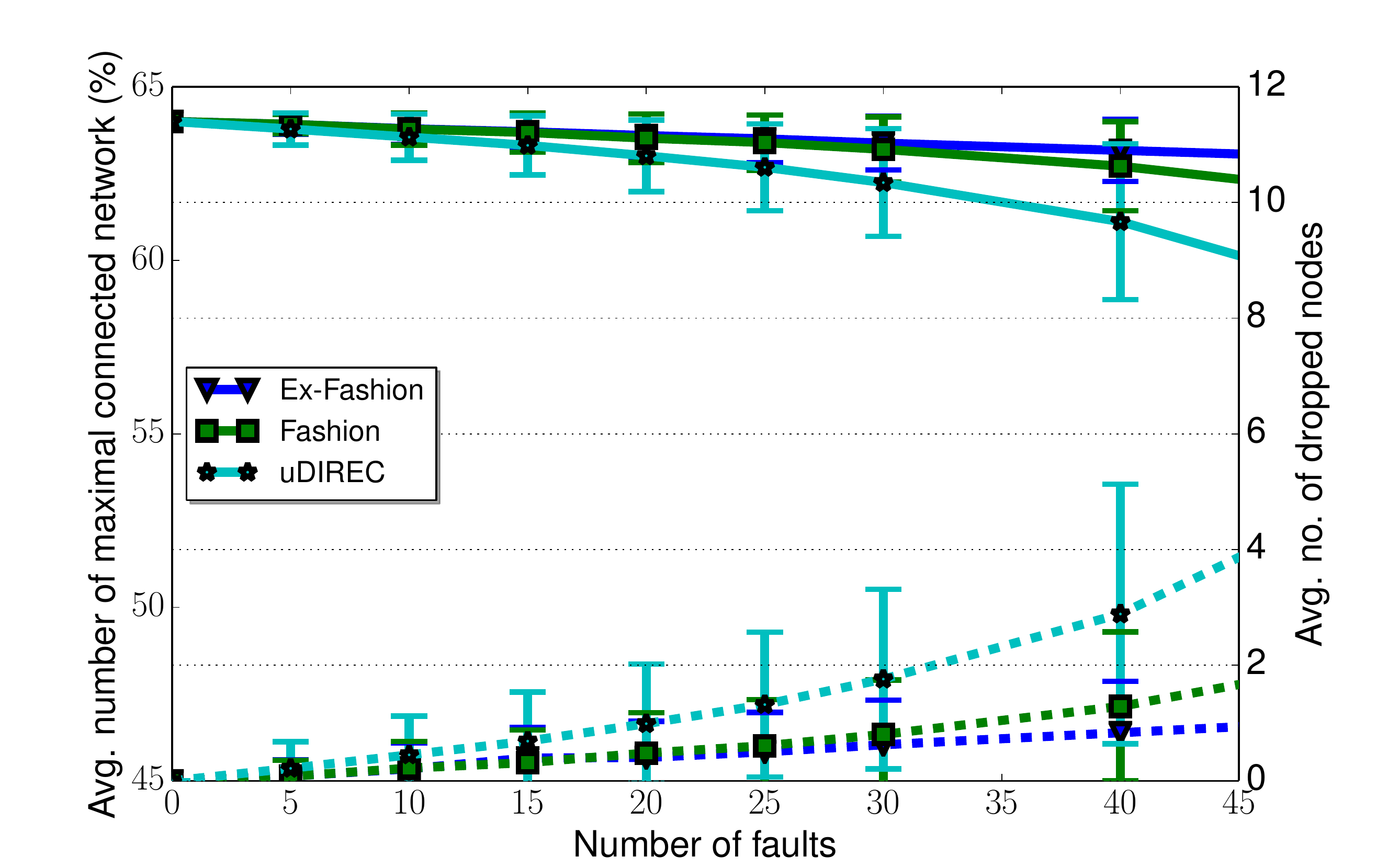}
\caption{Relationship of number of permanent faults and \textit{maximal connected network} and dropped nodes for different algorithms. 
Dotted curves are dropped nodes.}
\label{fig:fashion_vs_uDIREC}
\vspace{-0.2in}
\end{center}
\end{figure}

\begin{figure}[t]
\begin{center}
\includegraphics[width=0.4\textwidth]{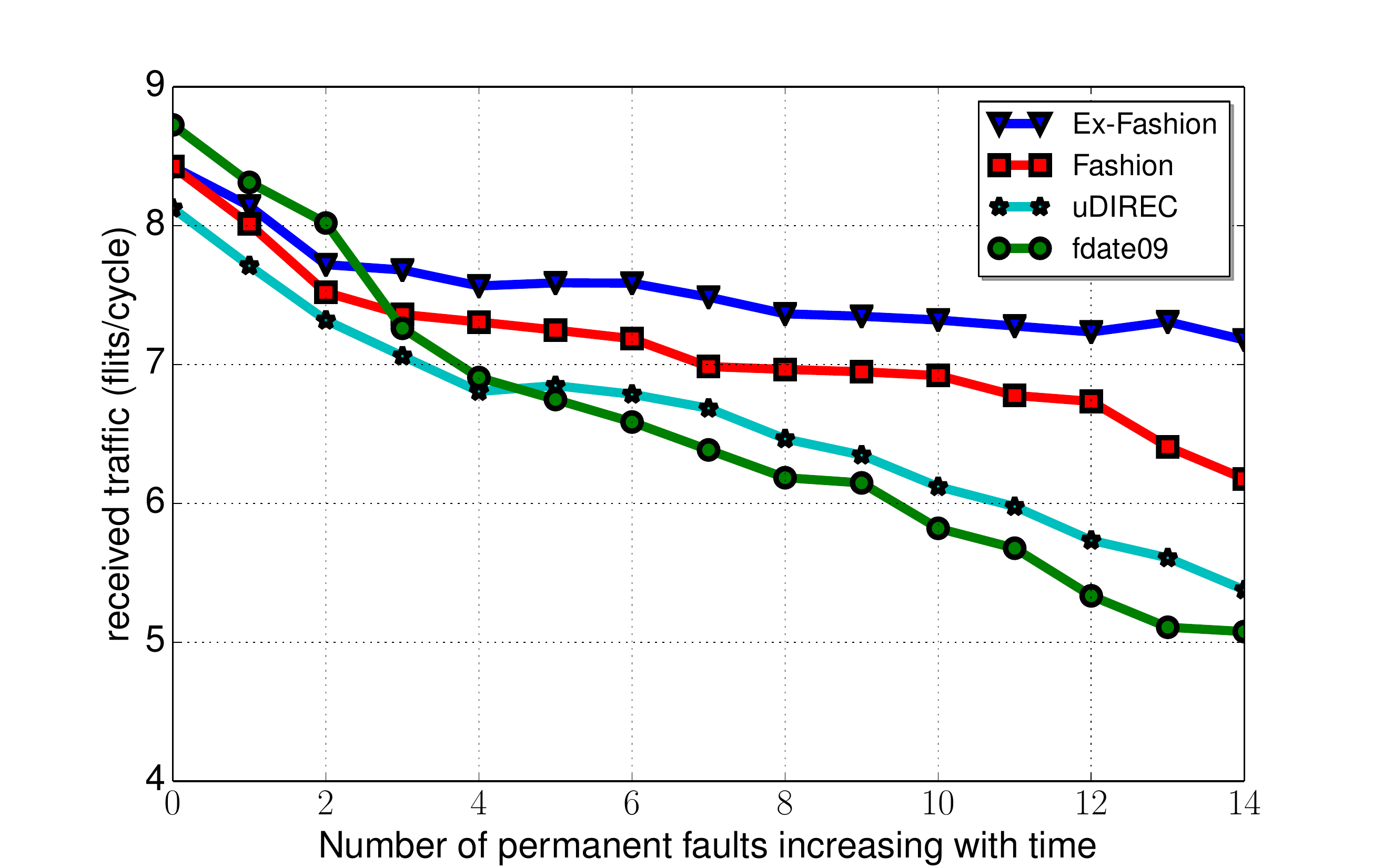}
\caption{Throughput degradation with increasing of number of permanent faults.}
\label{fig:throughput_faults}
\vspace{-0.2in}
\end{center}
\end{figure}

The uDIREC design only works when the underlying topology has two-way connectivity, a prerequisite that is not always present. 
We compare the average \textit{maximal connected network} percentage and the average number of disabled faultless nodes using uDIREC\cite{parikh2013udirec} with the same number of faults in the 8x8 2D mesh. As shown in Figure~\ref{fig:fashion_vs_uDIREC}, uDIREC dropped an average of 1.76 and 2.89 flawless nodes with standard deviation of 1.55 and 2.25, when there are 30 and 40 faults. 
In \textbf{\textit{Fashion}}, there are 54.3\% and 55.4\% fewer dropped nodes with 30 and 40 faults, resulting in a higher \textit{maximal connected network} percentage than uDIREC. In \textbf{\textit{Ex-Fashion}}, there are 64.5\% and 71.1\% fewer dropped nodes than uDIREC with 30 and 40 faults. 
Further, we compared NoC performance degradation with time-dependent components defects. For these experiments, we still assumed uniform-random fault distribution, and increased the number of new permanent faults over time. Figure~\ref{fig:throughput_faults} shows average throughput results of 100,000 simulations using \textsc{uniform-random} traffic pattern. \textbf{\textit{Fashion}} achieves 13.07\% and 18.9\% more throughput than uDIREC and \textit{F(date09)} when there are 10 faults, whereas \textbf{\textit{Ex-Fashion}} achieves 19.6\% and 25.77\% more throughput in the same situations. 

Since uDIREC applies Up*/Down* routing and has a relatively larger number of forbidden turns, it sacrifices routing 
flexibility which influences the connectivity and performance of the network. The average percentage of forbidden turns in uDIREC and \textbf{\textit{Fashion}} are 20.563\% and 17.665\% for 8x8 2D-meshes, results are shown in Table~\ref{tab:fashion_vs_uDIREC}.
The connectivity degree of \textbf{\textit{Ex-Fashion}} is 1.167x and 3.77x higher than \textbf{\textit{Fashion}} with 30 and 60 faults respectively, see Table~\ref{tab:statistic}. This is because the improved \textbf{\textit{Ex-Fashion}} architecture can better mitigate the negative effects of single link and virtual buffer faults. The study demonstrates a significant advantage by applying bidirectional link and unified virtual channel structure to \textbf{\textit{Fashion}}. The results of \textit{F(date09)} are not shown in Figure~\ref{fig:fashion_vs_uDIREC}, because a large portion of nodes are dropped and the \textit{maximal connected network} are dramatically decreased when the number of faults exceeds 30.

\begin{table}[h]
\begin{center}
\begin{tabular}{|c|c|c|c|c|} \hline\hline
{} &\multicolumn{2}{c}{8x8 2D mesh}&\multicolumn{2}{|c|}{16x16 2D mesh}\\
\cline{2-5}
{Fault num}&{uDIREC}&{Fashion}&{uDIREC}&{Fashion}\\ \hline
{10}& {20.510\%}&{19.798\%}&{20.148\%}&{20.000\%}\\ \hline
{20}& {20.702\%}&{19.222\%}&{20.293\%}&{19.954\%}\\ \hline
{30}& {20.780\%}&{18.374\%}&{20.403\%}&{19.855\%}\\ \hline
{40}& {20.677\%}&{17.458\%}&{20.500\%}&{19.842\%}\\ \hline
{50}& {20.499\%}&{16.705\%}&{20.597\%}&{19.741\%}\\ \hline
{60}& {20.209\%}&{14.434\%}&{20.698\%}&{19.721\%}\\ \hline\hline
\end{tabular}
\end{center}
\caption{Probability of forbidden turns}
\label{tab:fashion_vs_uDIREC}
\vspace{-0.2in}
\end{table}

\section{Conclusions}
\label{sec:concl}
Current and future technology scaling effects, in particular the decreasing transistor dependability, command more 
 effective resiliency-aware approaches to multicore and many-core systems designs. 
In this work, we propse and design a scalable, distributed and self-healing intelligent NoC router with minimal hardware overhead, named \textsc{\textit{Fashion}}. The new router architecture is a self-monitoring and self-reconfiguring design that allows the on-chip network to dynamically recover from permanent component failures. The router has bidirectional links, unified virtual channel structures and distributed intelligence modules for 
detecting component failures and self-adjusts to mitigate their negative system performance effects. 
To determine the importance of a defective link or router on the network connectivity, we adopt a distributed \textit{depth 
first search} algorithm capable of classifying faulty components as \textit{cut elements}. 
A \textit{node-table} based 
reconfiguration procedure is performed to provide deadlock-free and routing connectivity guarantees to the \textit{maximal connected graph} of the new network topology.  
The \textbf{\textit{Fashion}} router has a fully distributed operation mode with relatively constant on-chip 
overhead, which makes it more suitable for on-line large-scaled on-chip communication. 

\textbf{\textit{Fashion}}  can also be applied in the NoC Power-gating domain to provide deadlock-free paths to the NoC traffic in dynamically changing irregular topologies. With chips becoming increasingly more power-constrained today, there is a need for intelligent NoC power-gating schemes that can reduce static power consumption and at the same time provide deadlock-free low latency paths to NoC traffic.

In this work, we mainly focused on the microarchitecture layer of the on-chip network. In future work, we will examine in detail how the \textbf{\textit{Fashion}} router can interact with the operating system for runtime execution roll-backs, task placements, memory management and on-chip communication policies to provide a comprehensive fault-tolerant and fault-recovery mechanism to the computing system.




\end{document}